\documentclass[twocolumn,prl,aps,
showpacs,preprintnumbers,
superscriptaddress,
longbibliography]{revtex4-1}
\usepackage{graphicx}

\usepackage[pdf]{pstricks}

\usepackage{dcolumn}
\usepackage{bm}
\usepackage{color}
\usepackage{amsmath}
\usepackage{amssymb}
\usepackage{amsfonts}
\usepackage{esint}
\usepackage{hyperref}
\usepackage{braket}

\usepackage[position=top,caption=false]{subfig}
\usepackage{floatrow}
\usepackage{bbold}

\begin{document}

\title{Measurement of the topological Chern number by continuous probing of a qubit subject to a slowly varying Hamiltonian}

\author{Peng Xu}\email{xupengqh201461@163.com}
\affiliation{National Laboratory of Solid State Microstructures, School of Physics, Nanjing University, Nanjing 230039, China} \affiliation{Department of Physics and Astronomy, Aarhus University, 8000 Aarhus C, Denmark}

\author{Alexander Holm Kiilerich}
\affiliation{Department of Physics and Astronomy, Aarhus University, 8000 Aarhus C, Denmark}

\author{Ralf Blattmann}
\affiliation{Department of Physics and Astronomy, Aarhus University, 8000 Aarhus C, Denmark}

\author{Yang Yu}
\affiliation{National Laboratory of Solid State Microstructures, School of Physics,
Nanjing University, Nanjing 230039, China} \affiliation{Synergetic Innovation Center of Quantum Information and Quantum Physics, University of Science and Technology of China, Hefei 230026, China}

\author{Shi-Liang Zhu}
\affiliation{National Laboratory of Solid State Microstructures, School of Physics,
Nanjing University, Nanjing 230039, China} \affiliation{Synergetic Innovation Center of Quantum Information and Quantum Physics, University of Science and Technology of China, Hefei 230026, China}

\author{Klaus M\o lmer}\email{moelmer@phys.au.dk}
\affiliation{Department of Physics and Astronomy, Aarhus University, 8000 Aarhus C, Denmark}

\date{\today}

\begin{abstract}
We analyze a measurement scheme that allows determination of the Berry curvature and the topological Chern number
of a Hamiltonian with parameters exploring a two-dimensional closed manifold. Our method uses continuous monitoring of the gradient of the Hamiltonian with respect to one parameter during a quasiadiabatic quench of the other. Measurement backaction leads to disturbance of the system dynamics, but we show that this can be compensated by a feedback Hamiltonian. As an example, we analyze the implementation with a superconducting qubit subject to time-varying, near-resonant microwave fields， equivalent to a spin-1/2 particle in a magnetic field.

\end{abstract}

\maketitle

Since the pioneering work on topological phases of matter by Kosterlitz, Thouless and Haldane \cite{Kosterlitz1973,Thouless1982,Haldane1983}, quantum states with nontrivial topological properties have constituted a very active research field. Topological invariants, such as the Chern number \cite{QNiu1985}， characterize how the system ground state (or an excited state) varies when the system Hamiltonian explores a closed two-dimensional manifold of parameter values.
States with nontrivial topology in extended quantum systems are associated with, e.g., the variation of Bloch wave functions over a two-dimensional Brillouin zone \cite{Fruchart2013}, while the ground state of a single spin under variation of the direction of an external magnetic field constitutes a simple example where a topological transition may be witnessed by a discrete change in the Chern number \cite{Gritsev2012}.

Topological properties are not derived from a single state but depend on how, the eigenstates of different, continuously varied Hamiltonians are connected.  This is quantified by the Berry phase \cite{MVBerry1985}, associated with the evolution of a quantum system under adiabatic variation of the Hamiltonian. In addition to a dynamical phase, $-\int dt' E_n(t')/\hbar$, governed by the energy eigenvalue, the $n{th}$ eigenstate of the system acquires a geometric (Berry) phase, $i\int \langle \phi_n|\nabla_R |\phi_n\rangle\,dR$, along the curve explored by the Hamiltonian in parameter space. By parametrizing $R=(\mu,\nu)$, one can define the Berry connection
\begin{equation}
A_{\mu}=\left\langle\phi_n\right|i\partial_{\mu}\left|\phi_n\right\rangle,
\end{equation}
and the Berry curvature,
\begin{equation}
F_{\mu\nu}=\partial_{\mu}A_{\nu}-\partial_{\nu}A_{\mu},
\end{equation}
whose integral over any surface in parameter space by Stokes theorem yields the Berry phase for the closed boundary curve. When the surface integral is extended to a closed manifold (no boundary), it must yield $2\pi$ times an integer $C$, which in the classical Gauss-Bonnet theorem \cite{SSChern1944} is identified with the genus of the manifold and which in the quantum case defines the topological Chern number,
\begin{equation}\label{eq:ChernDefinition}
C=\frac{1}{2\pi}\int_{S}dS_{\mu\nu}F_{\mu\nu},
\end{equation}
where $dS_{\mu\nu}$ is a surface element in the parameter space. The restriction of Chern numbers to discrete values is associated with the topological properties of the system and explains the robustness of physical phenomena such as the integer quantum Hall effect \cite{VKlitzing1986,Haldane1988,Zhang2005}.
The inherent robustness implies reduced sensitivity to small perturbations and decoherence, suggesting quantum systems with nontrivial topology as promising building blocks in a quantum computer \cite{Nayak2008}.

In this work, we focus on the Berry curvature and Chern number associated with a particular system eigenstate $\ket{\phi_0}$ \cite{foot}.
Due to the dependence of the eigenstates on a manifold of different Hamiltonians, the Chern number is not directly related to an experimental observable.
Recently, however, Polkovnikov and Gritsev \cite{Gritsev2012} have shown that the slowly quenched dynamics of a quantum system permits identification of the Berry curvature as the expectation value of the gradient of the Hamiltonian: If the quantum system is subject to a slow quench of the Hamiltonian parameter $\nu$ with a rate of change $v$, the operator $f_{\mu}\equiv-\partial_{\mu}H$ has the expectation value
\begin{equation}\label{linRes}
\left\langle f_{\mu}\right\rangle=\left\langle\phi_0\right|f_{\mu}\left|\phi_0\right\rangle + vF_{\mu\nu} + O(v^{2}).
\end{equation}
The Berry curvature and the Chern number can hence be inferred from an experiment by measuring the  physical observable $f_\mu$ and extracting its linear dependence on $v$.

The protocol suggested in Ref. \cite{Gritsev2012} has recently been demonstrated by the evolution of a superconducting qubit subject to a time-dependent microwave field \cite{MDSchroer2014} whose detuning and Rabi frequency act in a manner equivalent to $z$ and $x$ magnetic field components on a spin-$1/2$ particle. Perfect adiabatic evolution accomplishes a Bloch vector rotation in the $xz$ plane. The gradient of the Hamiltonian along the azimuthal direction is then proportional to the spin component $\sigma_y$ which, indeed, acquires a finite expectation value proportional to $v$ under a quasiadiabatic sweep. The Bloch spheres inserted in Fig.~1 show the trajectories traced by the qubit Bloch vector for different quench times and for sweeps of the field parameters causing a flip of the spin (left insert) and a return to the initial state (right insert), respectively. The shorter quench times yield the larger nonadiabatic deviation of the quantum states and the larger expectation values of $\sigma_y$. Combining the outcomes of repeated projective measurements of $\sigma_y$ after different partial sweeps of the Hamiltonian parameters in Ref. \cite{MDSchroer2014} reveals a transition between different values of the Chern number depending on the directional solid angle explored by the effective magnetic field.

\begin{figure}
\centering
\includegraphics[width=1.0\columnwidth]{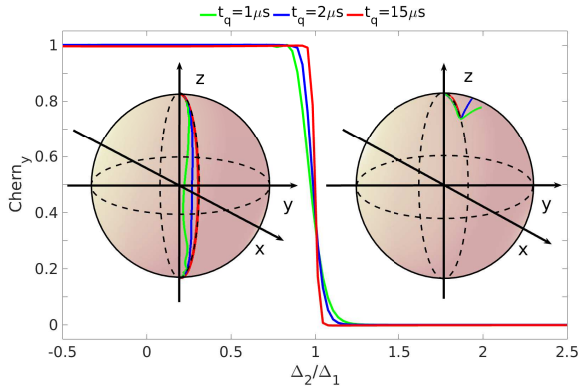}
\caption{(Color online) The plotted curves show the dependence  of the integral \eqref{eq:ChernDefinition} or \eqref{eq:chernform} on $\Delta_{2}/\Delta_1$ for different quench times $t_{q}=$ $1 \mu$s (green, lower curve), $2 \mu$s (blue, middle curve), and $15 \mu$s (red, upper curve). We assume $\phi=0$, $\Omega_{1}/2\pi=\frac{1}{3} \Delta_{1}/2\pi= 10$ MHz. The Bloch sphere on the left (right) shows the evolution of the quantum state during the quench for $\Delta_{2}=\frac{1}{3}\Delta_{1}$ ($\Delta_{2}=1.1\Delta_{1})$.
}
\label{F1DC}
\end{figure}
Rather than performing a sequence of partial sweeps, it is an attractive possibility to experimentally extract the Chern number and Berry phase by continuously monitoring the system during a full quench of the Hamiltonian. In this Rapid Communication, we shall analyze a protocol that allows such measurement.
As an alternative to projective measurements, we apply a weak dispersive probe which in the qubit example provides a signal with a mean value proportional to the time-dependent expectation value of $\sigma_y$. For sufficiently weak probing, we can disregard the measurement backaction on the system, but the signal will be dominated by detector shot noise. Stronger probing offers a better signal-to-noise ratio, but back-action will then modify the state and hence the future evolution of the system, reflecting the well-known and fundamental relationship between distinguishability and disturbance of quantum states by measurements \cite{CMCaves1980,AAClerk2010}.

To cancel the backaction of the strong continuous measurement, we propose to apply a simple continuous feedback on the qubit. As we are not aiming to determine an unknown quantum state but rather to probe a (classical) property of the Hamiltonian governing the system dynamics, there is no fundamental impediment against such a strategy, and we shall demonstrate its achievements with theoretical arguments and simulations.

As in Ref. \cite{MDSchroer2014}, we consider a qubit with two levels $\left|g\right\rangle$, $\left|e\right\rangle$ and transition frequency $\omega_{q}$.
In a frame rotating at the frequency of the applied microwave drive
$\omega_{d}$, the Hamiltonian can be expressed as ($\hbar = 1$)
\begin{equation}\label{eq:H}
H=\frac{1}{2}[\Delta\sigma_{z} + \Omega\sigma_{x}\cos\phi + \Omega\sigma_{y}\sin\phi].
\end{equation}
Here, $\sigma_{x}$, $\sigma_{y}$, and $\sigma_{z}$ are the Pauli operators, $\Delta=\omega_{q}-\omega_{d}$ is the detuning, $\Omega$ is the product of the microwave field amplitude and the qubit transition dipole moment, and $\phi$ is the phase of the microwave field. We parametrize the dynamical quench of the detuning and Rabi frequency by a time-dependent polar angle variable $\theta(t)=vt$,
\begin{equation}
\Delta=\Delta_{1}\cos\theta + \Delta_{2}, \qquad \Omega=\Omega_{1}\sin\theta.
\end{equation}
The quench speed $v=\pi/t_q$, appearing in Eq.~(\ref{linRes}), is determined by the quench time $t_q$.

The Bloch vector components $(x,y,z)$ of a two-level system are defined by the expansion of the density matrix on Pauli spin matrices,
\begin{align}\label{eq:bloch}
\rho = \frac{1}{2}\left(\mathbb{1}+x\sigma_x+y\sigma_y+z\sigma_z\right),
\end{align}
i.e., $u_\rho = \mathrm{Tr}(\sigma_u\rho_t)$ for $u =x,y,z$.
As shown by the left inset in Fig.~\ref{F1DC}, if $\Delta_{2}/\Delta_{1} < 1$ (assuming both are positive), the microwave frequency performs a chirp across the qubit resonance and the adiabatic qubit Bloch vector $(\Omega\cos\phi, \Omega\sin\phi,\Delta)/\sqrt{\Omega^2+\Delta^2}$ passes from the north to the south pole, while if $\Delta_{2}/\Delta_{1}> 1$, both the initial and final states of the adiabatic evolution are represented by the north pole of the Bloch sphere (right inset).

 A rapid quench of the parameter $\theta$, i.e., a small value of $t_q$, causes deviations from the adiabatically evolved state, as clearly illustrated in Fig.~1. Equations~(\ref{eq:ChernDefinition}) and (\ref{linRes}) now yield the expression for the Chern number,
\begin{equation}\label{eq:chernform}
C_y=-\int_{0}^{\pi}\frac{\Omega_{1}}{2v}\left\langle\sigma_{y}\right\rangle \sin\theta d{\theta},
\end{equation}
where we have used the independence of the Berry connection and Berry curvature on the azimuthal angle $\phi$ to reduce Eq.~(\ref{eq:ChernDefinition}) to a one-dimensional integral along the $\phi=0$ longitude.

\begin{figure*}
\begin{center}
\subfloat[][\label{subfig:a}]{\includegraphics[width=0.29\columnwidth]{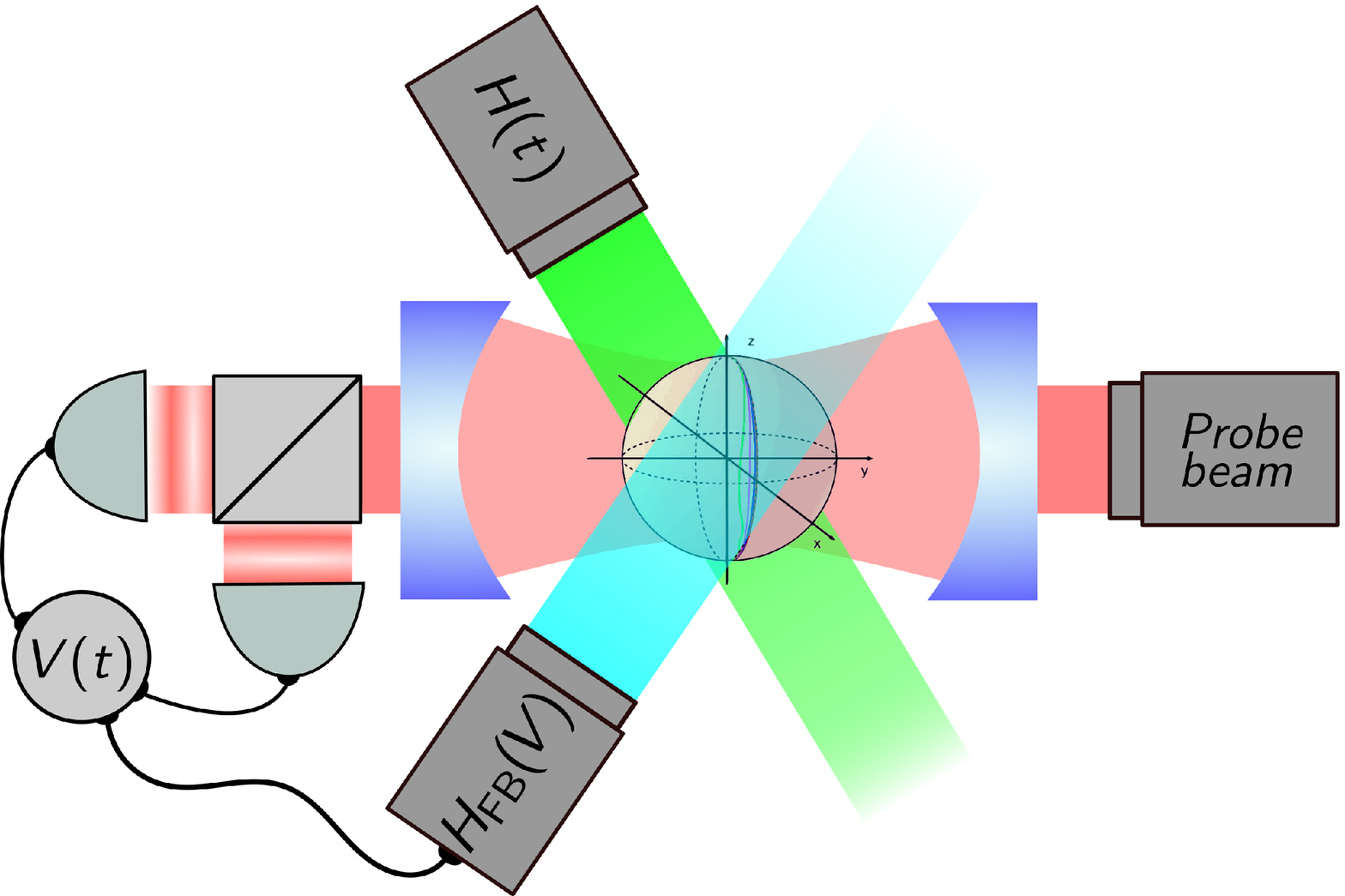}}
\subfloat[][]{\includegraphics[width=0.31\columnwidth]{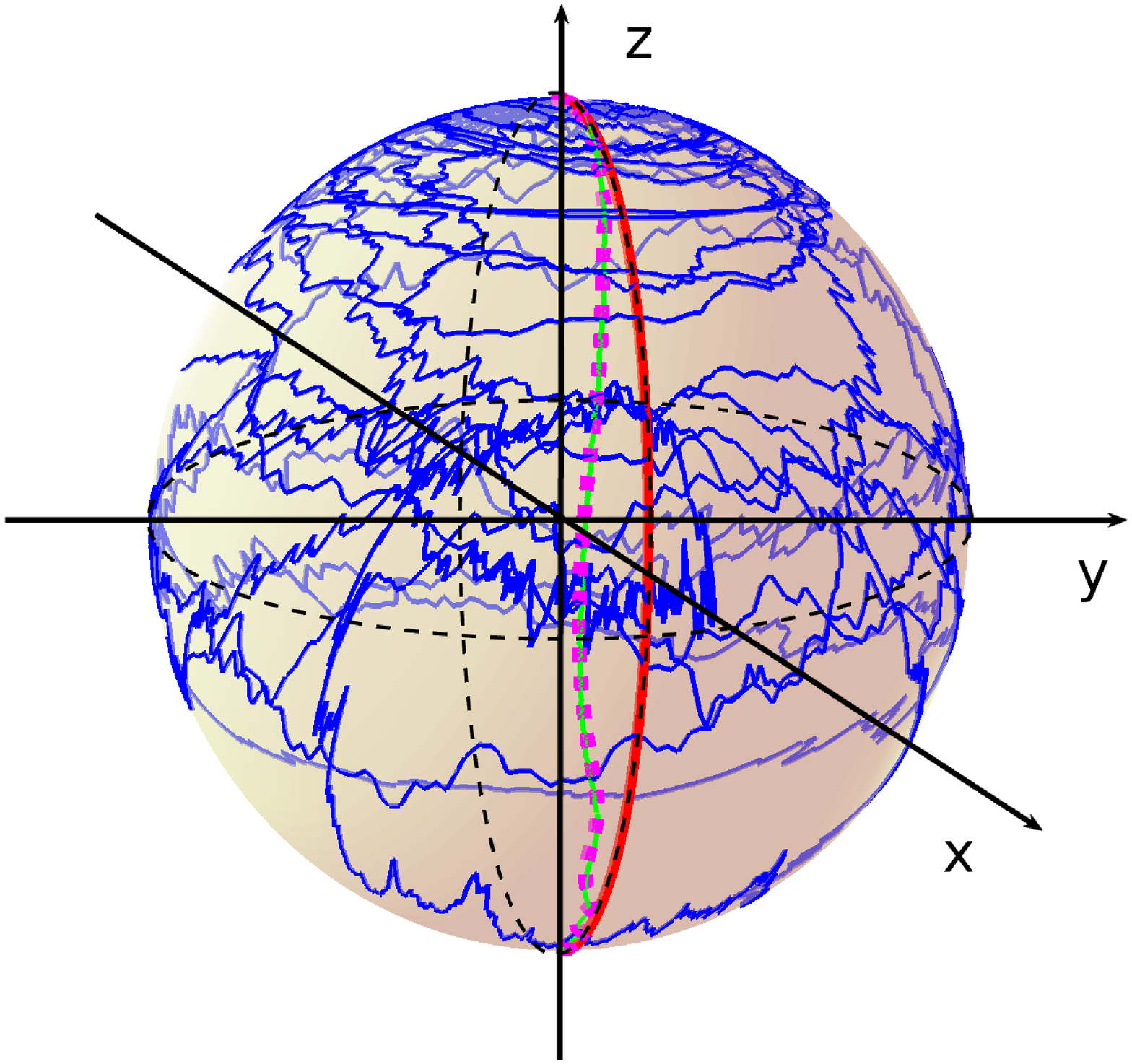}
\label{subfig:b}}
\subfloat[]{\includegraphics[clip,trim={0cm 0 0 0},width=0.34\columnwidth]{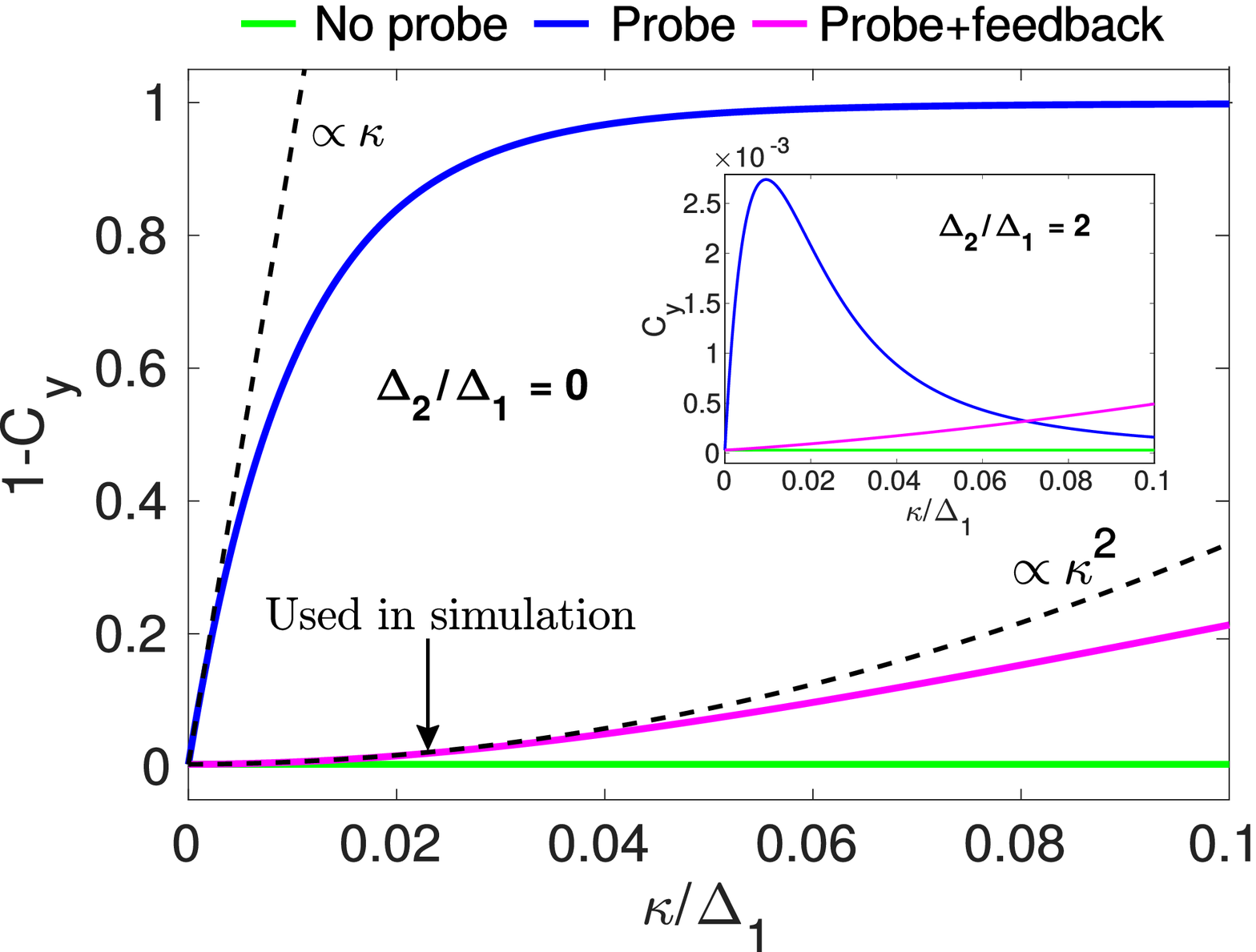}
\label{subfig:c}}
\
\subfloat[]{\includegraphics[clip,trim={0cm 0 0 0},width=0.32\columnwidth]{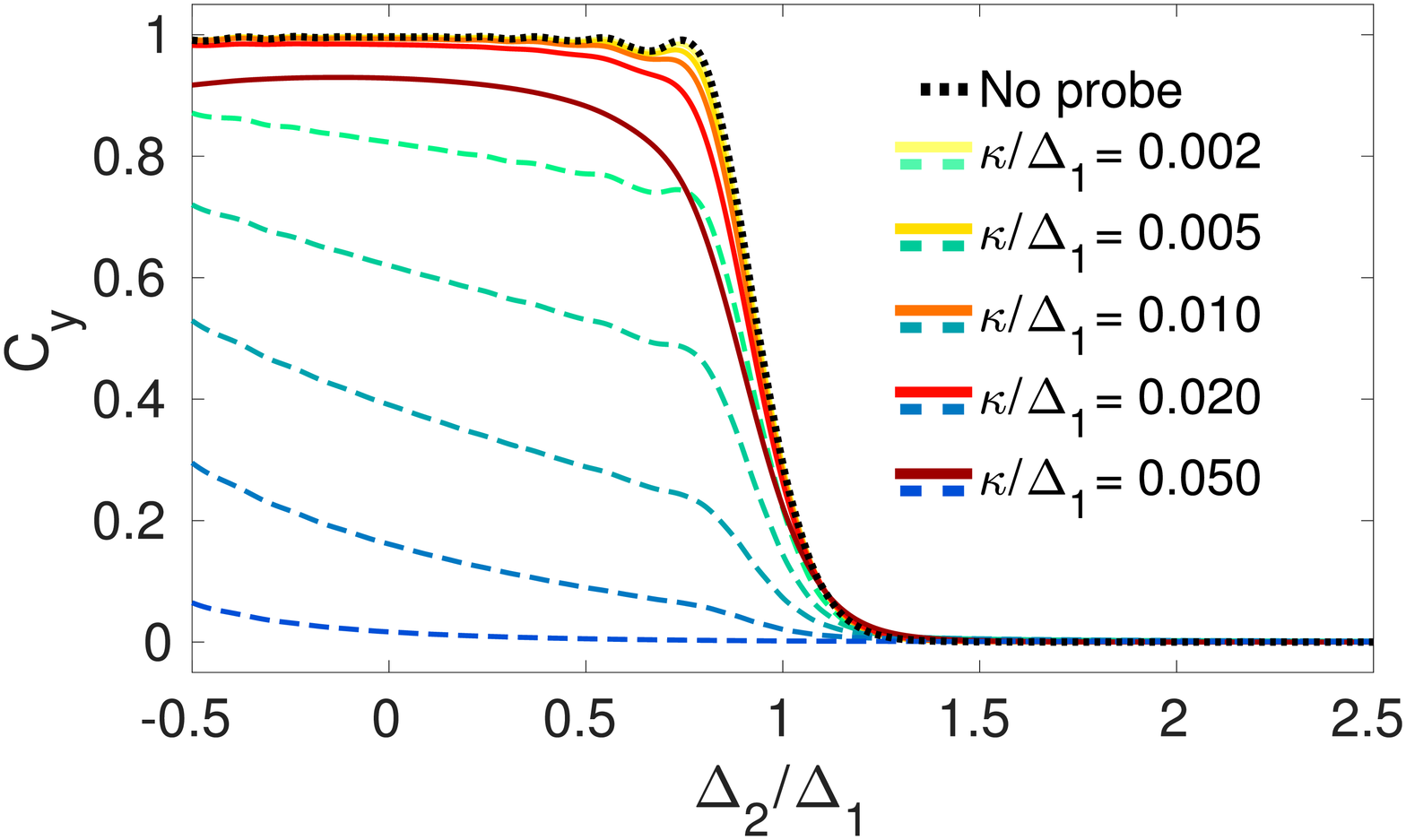}
\label{subfig:d}}
\subfloat[]{\includegraphics[clip,trim={0cm 0 0 0},width=0.64\columnwidth]{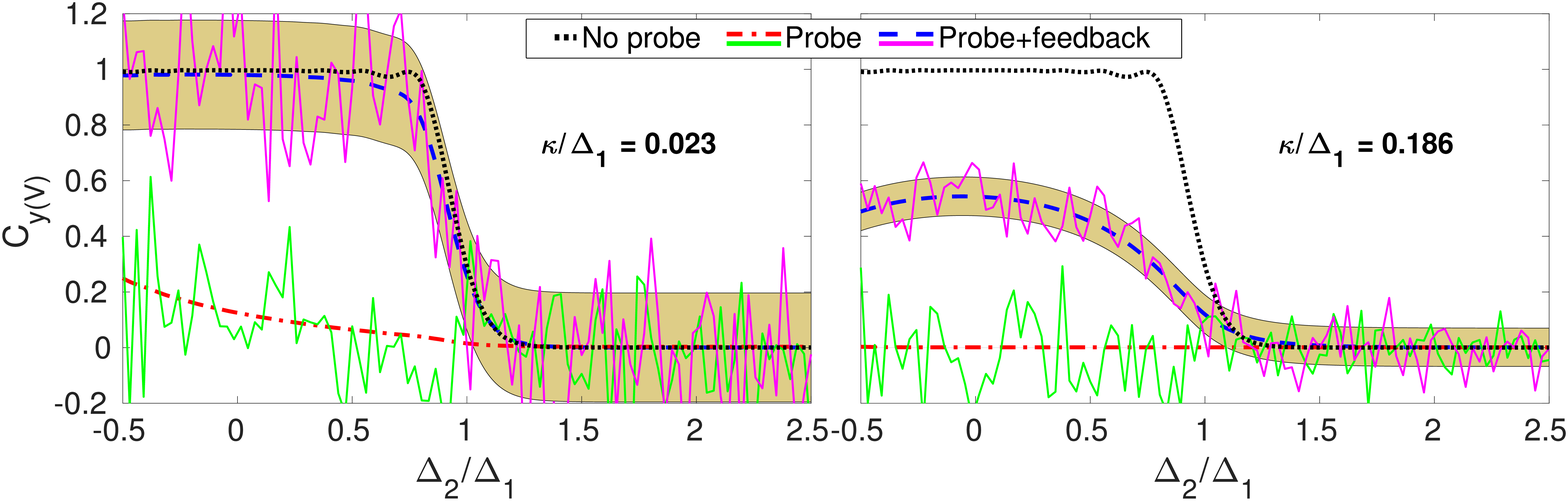}
\label{subfig:e}}
\end{center}
\caption{(Color online) (a) A two-level system  driven by a time-dependent Hamiltonian is monitored by homodyne detection of a transmitted microwave field. The stochastic homodyne current can be used to control a feedback microwave field applied to the qubit. (b) The dynamical evolution of the quantum state is calculated with $\Omega_{1}=\Delta_{2}=\frac{1}{3}\Delta_{1} = 10$ MHz.
Results are shown for adiabatic evolution with $t_{q}=15\,\mu$s (red, in $xz$ plane) and for quasiadiabatic evolution with $t_{q}=1\,\mu$s (green, blue, and dotted magenta).
The dotted magenta curve depicts the evolution absent probing, the blue curve shows the evolution with high measurement strength ($\kappa/2\pi = 0.37$ MHz), and the green curve (matching the dotted magenta curve) indicates the evolution under a strong measurement employing the feedback described in the text.
(c) The deviation in the Chern number (integral in Eq.~(\ref{eq:chernform}) calculated from the unconditional evolution) from the expected integer value, introduced by measurement backaction. Curves are displayed as a function of the probing strength $\kappa$ for the cases of no probing (green, lower line), probing (blue, upper line), and probing with feedback (magenta, middle line). Results are shown for $\Delta_2=0$, where in the absence of probing, $C_y=1$, and the inset refers to $\Delta_2=2\Delta_1$, where we expect $C_y = 0$.
(d) The value of $C_y$ for different detuning ratios $\Delta_{2}/\Delta_1$ and different values of the probing strength. We use $\Omega_1 = \frac{1}{3}\Delta_1$ and the remaining parameters are set to optimize the measurement proposal (see text); $\Delta_1/2\pi = 16.1$ MHz and $t_q = 0.96\,\mu$s. Lower, dashed curves (cool colors) depict the case of probing without feedback and the upper, full curves (warm colors) show the case with feedback. The dotted black curve shows $C_y$ without probing.
(e) The Chern number for different detuning ratios $\Delta_{2}/\Delta_1$ with the optimized measurement strength
$\kappa/2\pi = 0.37$ MHz (left panel) and with a much larger strength $\kappa/2\pi = 3.0$ MHz (right panel). Shaded areas illustrate error bars with $N=383$ experimental repetitions.
Noisy, magenta (upper) and green (lower) curves, consistent with the error bars and ensemble average signals $C_y$, illustrate the results for $C_V$ from simulations of the measurement and feedback procedure.
\label{F2abc-(upper)-(lower)}}
\end{figure*}

A continuous measurement of the $\sigma_y$ observable is performed by injecting a probe field which interacts dispersively with the qubit, and the reflected signal is amplified and demodulated to yield an output voltage signal $V(t)$ [Fig.~\ref{subfig:a}]; see, e.g., Refs. \cite{KWMurch2013,MHatridge2013,SJWeber2014}. In dimensionless units, $V(t)$ has the form
\begin{align}\label{eq:J}
V(t) = \mathrm{Tr}(\sigma_y\rho_t) + \frac{d W_t}{2\sqrt{\eta\kappa}dt},
\end{align}
with a mean value given by the desired qubit expectation value and with random, shot-noise fluctuations. The infinitesimal white-noise Wiener increment $d W_t$ has zero mean and variance $dt$ \cite{KJacobsStochastic2010}. The noise term dominates the readout if the probe field strength $\kappa$ is weak or if the detector efficiency $\eta$ is low.

The Chern number $C_V$ is deduced from the probe signal by performing the integral (\ref{eq:chernform}) with the signal $V(t)$ in place of the expectation value $\langle \sigma_y \rangle$. Due to the Gaussian fluctuations in the signal, the estimated Chern number samples a normal distribution with mean $C_y$, while the white-noise component in the signal \eqref{eq:J} yields a statistical error in the estimation of $C_y$ from a single quench experiment of magnitude
\begin{align}\label{eq:ChernError}
\Delta C_V^{(1)}= \Omega_1\sqrt{\frac{t_q}{32\eta\kappa}}.
\end{align}
Upon $N$ repetitions of the experimental sequence, this is reduced to $\Delta C_V^{(N)}=  \Delta C_V^{(1)}/\sqrt{N}$.
As we need the uncertainty to be well below unity to discern the integer values of $C_y$, we demand $\sqrt{\eta\kappa v N} \simeq \Omega_1$ which for small $N$ brings us in conflict with either the near adiabaticity or the assumption of negligible measurement backaction.

To account for the measurement backaction related to continuous monitoring of the $\sigma_y$ observable, we have recourse to the quantum theory of measurement in which the \emph{conditional} evolution of the density matrix $\rho_t$ of the probed qubit is governed by the stochastic master equation (SME) \cite{KJacobs2006,HMWiseman2009},
\begin{align}\label{eq:SME}
d\rho_t &= -i[H(t),\rho_t] dt+ \kappa\mathcal{D}[\sigma_y]\rho_t dt + \sqrt{\eta\kappa}\mathcal{H}\left[\sigma_y\right]\rho_t dW_t.
\end{align}
The first term represents the evolution absent probing as described by the Schr\"odinger equation and depicted in the insets of Fig.~\ref{F1DC}.
Because of the interaction with the probe field, this evolution is supplemented by the latter two terms, where we define the superoperators
\begin{align}
\mathcal{D}[a]\rho &=a\rho a^\dagger -\frac{1}{2}\left\{a^\dagger a,\rho\right\},
\\
\mathcal{H}[a]\rho &=a\rho+\rho a^\dagger -\mathrm{Tr}(a\rho+\rho a^\dagger) \rho,
\end{align}
responsible for deterministic decoherence and stochastic measurement backaction, respectively.

Under perfect detection ($\eta=1$), the conditional state remains pure and the system lives on the surface of the sphere, while with imperfect detection ($0\leq\eta<1$) our inability to
trace the state leads to a mixed state \textit{inside} the Bloch sphere. Henriet  \emph{et al.} \cite{LHenriet2016} have investigated the behavior of the mixed state Bloch vector components and the resulting value of the integral (\ref{eq:chernform}) in the case of a qubit coupled to a reservoir of bath degrees of freedom. That study, for not too strong bath coupling, justifies the determination of the Chern number by our procedure, and for clarity we restrict our attention to the $\eta=1$ case. See Ref. \cite{Supplemental} for a discussion of the implications of finite detection efficiency.

In the case of a strong probing, the signal will be less noisy but the system will be attracted to one of the eigenstates of the monitored observable, $\sigma_y$.
At this point, both of the last terms in \eqref{eq:SME} vanish;
i.e., in the limit of very large $\kappa$, the theory describes a projective measurement.
In this case, as exemplified by the simulated blue curve in Fig.~\ref{subfig:b},
the evolution deviates strongly from the unprobed dynamics and integration of the measurement signal will not yield the Chern number.

This fact can be addressed further by considering the estimate for the Chern number obtained by averaging over a large number $N\gg 1$ of experimental runs.
The ensemble average or \textit{unconditional} evolution of the system follows from the master equation~(\ref{eq:SME}), but without the last term, which because of the white-noise increment averages to zero.
The second term describes the average measurement-induced decoherence, which introduces an error in the Chern number as defined in Eq.~(\ref{eq:chernform}).
The dashed lower curves in Fig.~\ref{subfig:d} show the Chern number calculated from the solution of the unconditional master equation during a single sweep of the parameter $\theta$. Results are shown for a range of detunings and we assume different values of the measurement strength. The figure confirms our expectations: For (very) weak probing, the procedure yields a result which is compatible with the correct value of the Chern number, but by Eq.~(\ref{eq:ChernError}) distinguishing this value requires a large number of experimental repetitions, $N\gtrsim \Omega_1^2/\eta\kappa v$. For stronger probing, the noise is reduced, but the average measurement backaction disturbs the evolution and we do not obtain a good candidate value for the Chern number.
This conclusion is supported by the blue upper curve in Fig.~\ref{subfig:c}, showing the error in the Chern number at $\Delta_2/\Delta_1=0$ as a function of the probe strength $\kappa$. For small $\kappa$, the error is $\propto \kappa$, reflecting a system evolution which is maintained only when subject to very weak probing.
As seen from the inset, at $\Delta_2/\Delta_1 = 2$ where the Chern number should be $0$, the effect of the probing is three orders of magnitude lower.


\floatsetup[figure]{style=plain,subcapbesideposition=top}
Since the statistical error bars are reduced by a factor $1/\sqrt{N}$ if the quench protocol is repeated $N$ times, we can obtain the Chern number by repeated probing with a weak probing strength, but we shall now turn to an alternative approach depicted schematically in Fig. \ref{subfig:a}.
As the SME (\ref{eq:SME}) provides the evolution of the qubit state, conditioned on the measurement signal, if the state remains pure (which is the case if the measurement is performed with unit efficiency), we can compensate for the measurement back-action by a unitary qubit rotation.

Such an approach has been proposed \cite{HFHofmann1998,JWang2001} and recently applied in experiments to stabilize an arbitrary state \cite{PCampagne-Ibarcq2016} and to sustain Rabi oscillations \cite{RVijay2012} of a superconducting qubit long after the system would have decayed by spontaneous emission.
These works apply feedback based on quadrature detection of the resonance fluorescence from a two-level system.
Here we derive a similar capability for a dispersive readout of the $\sigma_y$ component, and we show that it can be used to measure an unknown small deviation of the quantum state from the adiabatic evolution without severely affecting the state.

Figure~\ref{subfig:a} illustrates how the probe signal is directly applied to control the drive strength and detuning parameters of a feedback microwave signal, acting on the qubit by the Hamiltonian,
\begin{align}\label{eq:Hv}
H_\mathrm{V}(t) &= \kappa(\alpha \sigma_x+\beta\sigma_z)V(t).
\end{align}
As detailed in the Supplemental Material \cite{Supplemental}, if we assume the values
\begin{align} \label{eq:feedbackParameters1}
\begin{split}
\alpha = 2z, \qquad
\beta = -2x,
\end{split}
\end{align}
all the stochastic terms, linear in $dW_t$, cancel. There are, however, corrections which are linear in $dt$ due to correlations between the $dW_t$ terms in the SME and in the feedback. Adding a second state-dependent Hamiltonian term,
\begin{align}\label{eq:Hab}
H_\mathrm{D}(t) &= \kappa(a \sigma_x + b\sigma_z),
\end{align}
with \cite{Supplemental}
\begin{align}\label{eq:feedbackParameters2}
\begin{split}
a &= -yz, \qquad
b = xy,
\end{split}
\end{align}
ensures that the system evolves according to the quench Hamiltonian only.

While this feedback may perfectly cancel the measurement backaction, it assumes knowledge of the quantum state and hence of the very observable that we want to measure. However, since our scheme assumes only a small deviation of the quantum state from the analytically known eigenstate of the time-dependent Hamiltonian, we may approximate our feedback operation with the one that we would have applied to restore the adiabatic state if that had been subject to the measurement; i.e., we approximate the values $x,y,z$ in (\ref{eq:feedbackParameters1}) and (\ref{eq:feedbackParameters2}) by the analytically known (time-dependent) Bloch vector components $(\Omega,0,\Delta)/\sqrt{\Omega^2+\Delta^2}$, and obtain a recipe that may be applied in real time during an experiment.

In the case of a quasiadiabatic sweep, the state only deviates slightly from the adiabatic eigenstate, and the feedback error is expected to be small. To assess the effect of this discrepancy, we turn again to the average evolution over many experimental runs, now with the feedback turned on. The unconditional master equation for a system subject to the feedback, Eqs.~(\ref{eq:Hv}) and (\ref{eq:Hab}), is given in the Supplemental Material \cite{Supplemental}, and from the solution of this equation during a full sweep of $\theta$, Eq.~(\ref{eq:chernform}) yields a Chern number candidate. The purple (middle) curve in Fig.~\ref{subfig:c} shows that the feedback indeed effectively removes the effect of the backaction and that the slight deviation from the adiabatic state used in the feedback parameters yields an error in the Chern number scaling as $\propto \kappa^2$.
This allows much larger values of the measurement strength to be used and hence a significant noise reduction [cf. Eq.~(\ref{eq:ChernError})] without severely affecting the estimate. However, once the probing strength is increased beyond $\kappa\simeq 0.03\Delta_1$ the evolution is strongly affected by the feedback, which effectively pulls the state toward the adiabatic eigenstate for which the $\sigma_y$ component and hence $C_y$ is zero.
In other words, the imperfect feedback sets an upper limit to the probing strength and hence the achievable noise reduction.

In our choice of the experimental parameters $\Omega_1$, $t_q$, and $\kappa$, we are hence presented with a tradeoff, leading to a simple optimization problem: The noise Eq.~(\ref{eq:ChernError}) must be minimized under the constraint that the error in the mean Chern number $C_y$ calculated from Eq.~(\ref{eq:chernform}) is small. As an example, we here perform the optimization at a representative ratio of the detuning parameters here we use $\Delta_2=0$ and require $C_y>0.98$.
The resulting dependence of the integral (\ref{eq:chernform}) on the ratio of the detuning parameters is compared to the unprobed results by the upper curves in Fig.~\ref{subfig:d} and the optimal parameters are given in the figure caption. It is evident that by employing the feedback, we can use much stronger probing and still obtain the expected variation of the Chern number with the detuning parameter.

The achievable noise reduction subject to the constraint imposed by the error tolerance does not allow a full determination of the Chern number from a single physical measurement and one has recourse to perform a number of repetitions. If, for instance, one wishes to distinguish a Chern number of $C_V>0.98$ from $C_V=0$ to five-$\sigma$ accuracy (that is, a probability to assign a wrong Chern number of $0.00003\%$), one must require $\Delta C_V^{(N)}\leq 0.196$. This implies that $N\simeq 383$ repetitions are needed. The corresponding error bars are illustrated along with the mean value over a range of detuning parameters in the left panel of Fig.~\ref{subfig:e}.
For comparison, we show in the right panel also the value of $C_y$ for much stronger probing $\kappa/2\pi  = 3$ MHz with error bars corresponding to the same $N$. Notice how the mean Chern number is reduced from $>0.98$ to around $0.5$, but that the same number of repetitions produces much smaller error bars, according to Eq.~(\ref{eq:ChernError}).

We have implemented the feedback scheme in numerical simulations of the quench and measurement procedure by solving the feedback SME given in Ref. \cite{Supplemental}. In Fig.~\ref{subfig:b}, the dotted magenta curve shows how the evolution of the Bloch vector during a single quench is virtually indistinguishable from the green curve representing the evolution in the absence of measurements. This confirms that with these optimized parameters, the feedback succeeds in canceling the measurement backaction while maintaining a quasi-adiabatic evolution, allowing a dynamical phase transition to be observed.

In  Figs.~\ref{subfig:e}, we plot results for the Chern number $C_V$ obtained from the simulated measurement records. In the left panel, we use the optimized sweep parameters and measurement strength and in the right panel we use a much stronger measurement strength.
The lines are guides to the eye and their fluctuations are representative of the estimation error on the Chern number from the signal recorded in $N\simeq 383$ sweeps.
The simulations are consistent with the mean value obtained from the unconditional evolution and the shaded areas, illustrating the analytic noise estimate Eq.~(\ref{eq:ChernError}).

In summary, we have proposed to measure the topological Chern number by continuously probing a physical observable during a full quench of the system Hamiltonian. Our results show that relatively few experimental repetitions are required to obtain an acceptable signal-to-noise ratio.
While it may also be possible to employ interference protocols to determine geometric phases of a qubit degree of freedom \cite{TBitter1987,DSuter1988}, the arguments and more complicated examples offered in Ref. \cite{Gritsev2012} encourage efforts to experimentally extract the Berry curvature and Chern number by direct measurement of a physical observable. Whether more complex physical systems also permit an explicit feedback scheme and noninvasive measurements depends on the specific system and on practical experimental capabilities.

Let us finally comment that our continuous noninvasive measurement may be applied to evaluate the integrated effect of any weak perturbation acting on a system, e.g., a small variation in a Hamiltonian strength parameter. For many measurements, a longer probing time $T$ will yield an improved scaling of the resolution with $1/\sqrt{T}$. Such improvement does not occur for the measurement of the Chern number where the mean signal, $\left\langle\sigma_y\right\rangle$, vanishes, while the integrated noise increases in the adiabatic limit of long quench times.

\begin{acknowledgements}
Peng Xu acknowledges helpful discussions with C. Yu. This work was supported by the the NKRDP of China (Grant No. 2016YFA0301800).
Peng Xu was also supported by a fellowship from the China Scholarship Council.
A. H. K., R. B., and K. M. acknowledge financial support from the Villum Foundation.
A. H. K. further acknowledges support from the Danish Ministry of Higher  Education  and  Science.
\end{acknowledgements}


%

\pagebreak
\widetext
\begin{center}
\textbf{\large Supplemental Materials: Measurement of the topological Chern number by continuous probing of a qubit subject to a slowly varying Hamiltonian}
\end{center}

\setcounter{equation}{0}
\setcounter{figure}{0}
\setcounter{table}{0}
\setcounter{page}{1}
\makeatletter
\renewcommand{\theequation}{S\arabic{equation}}
\renewcommand{\thefigure}{S\arabic{figure}}
\renewcommand{\bibnumfmt}[1]{[S#1]}
\renewcommand{\citenumfont}[1]{S#1}

\section{Feedback master equation and optimal parameters}
\label{sec:AppendixA}

To describe the evolution of the density matrix of the system subject to measurement and feedback, we must apply It\^o-calculus to treat the noise component of $V(t)$ in the feedback $H_\mathrm{V}$ as it is correlated with the noise in the SME (10) of the main text and yields a $dW_t^2=dt$ first order contribution in $dt$.
Assuming no delay in the feedback channel and perfect detection ($\eta=1$), the stochastic master equation for a monitored system subject to continuous feedback has been derived by Wiseman and Milburn \cite{2009HMWiseman},
\begin{align}\label{eq:meFeedback}
\begin{split}
d\rho &= -i[H(t),\rho] dt-i[H_\mathrm{D}(t),\rho]dt -i[\tilde{H}_\mathrm{V}(t),\sigma_y\rho+\rho \sigma_y]dt
\\
&\quad+
\kappa\mathcal{D}[\sigma_y]\rho dt+(\kappa\eta)^{-1}\mathcal{D}[\tilde{H}_\mathrm{V}(t)]\rho dt
\\
&\quad+
\sqrt{\eta \kappa}\mathcal{H}[\sigma_y-i\tilde{H}_\mathrm{V}(t)/(\eta\kappa)]\rho dW_t,
\end{split}
\end{align}
where $\tilde{H}_\mathrm{V}(t) = H_\mathrm{V}(t)/V(t)$.

To derive the feedback parameters $\alpha$, $\beta$, $a$ and $b$ given in (14,16) in the main text, it is convenient to represent the density matrix in terms of the Bloch vector components as in Eq. (7) of the main text.
For $\eta=1$, the stochastic feedback master equation \eqref{eq:meFeedback} is equivalent to a set of stochastic Bloch equations,
\begin{align}\label{eq:rhoBloch}
\begin{split}
dx &=dx_\mathrm{unitary}+dx_{\mathrm{ba}}+dx_{\mathrm{fb}}
\\
dy &=dx_\mathrm{unitary}+dy_{\mathrm{ba}}+dy_{\mathrm{fb}}
\\
dz &= dx_\mathrm{unitary}+dy_{\mathrm{ba}}+dy_{\mathrm{fb}},
\end{split}
\end{align}
where the first terms describe unitary evolution governed by the Hamiltonian,
\begin{align}\label{eq:rhoBloch}
\begin{split}
dx_{\mathrm{ba}} &=-2\kappa x dt-2\sqrt{\kappa}xy dW_t
\\
dy_{\mathrm{ba}} &=2\sqrt{\kappa}(1-y^2) dW_t
\\
dz_{\mathrm{ba}} &= -2\kappa z dt-2\sqrt{\kappa}yz dW_t
\end{split}
\end{align}
give the measurement back-action in the original SME, and
\begin{align}\label{eq:rhoBloch}
\begin{split}
dx_{\mathrm{fb}} &=\kappa\left(-by-2\beta-\frac{\beta^2}{2}x+\frac{\alpha \beta}{2} z\right)dt-\sqrt{\kappa}\beta y dW_t
\\
dy_{\mathrm{fb}} &=\kappa\left(-az+bx-\frac{(\alpha^2+\beta^2)}{2}y\right)dt+\sqrt{\kappa}(\beta x-\alpha z) dW_t
\\
dz_{\mathrm{fb}} &= \kappa\left(ay+2\alpha-\frac{\alpha^2}{2}z+\frac{\alpha \beta}{2}x\right)dt+\sqrt{\kappa}\alpha y dW_t
\end{split}
\end{align}
incorporate the extra terms associated with the feedback.
The purpose of the feedback is to cancel both the deterministic and the stochastic effects of the probing. This requirement ($0 = du_{\mathrm{ba}}+du_{\mathrm{fb}}$ for $u =x,y,z$) can be met only for a pure state, obeying $1=x^2+y^2+z^2$, and we readily identify how the feedback parameters should depend on the Bloch vector of the current state of the system,
\begin{align}\label{eq:feedbackParameters}
\begin{split}
\alpha &= 2z,
\\
\beta &= -2x,
\\
a &= -yz, \quad
\\
b &= xy.
\end{split}
\end{align}
These are the control parameters given in (14,~16) of the main text.

\begin{figure}[h!]
\centering
\subfloat[][\label{subfig:3a}]{\includegraphics[width=0.45\textwidth]{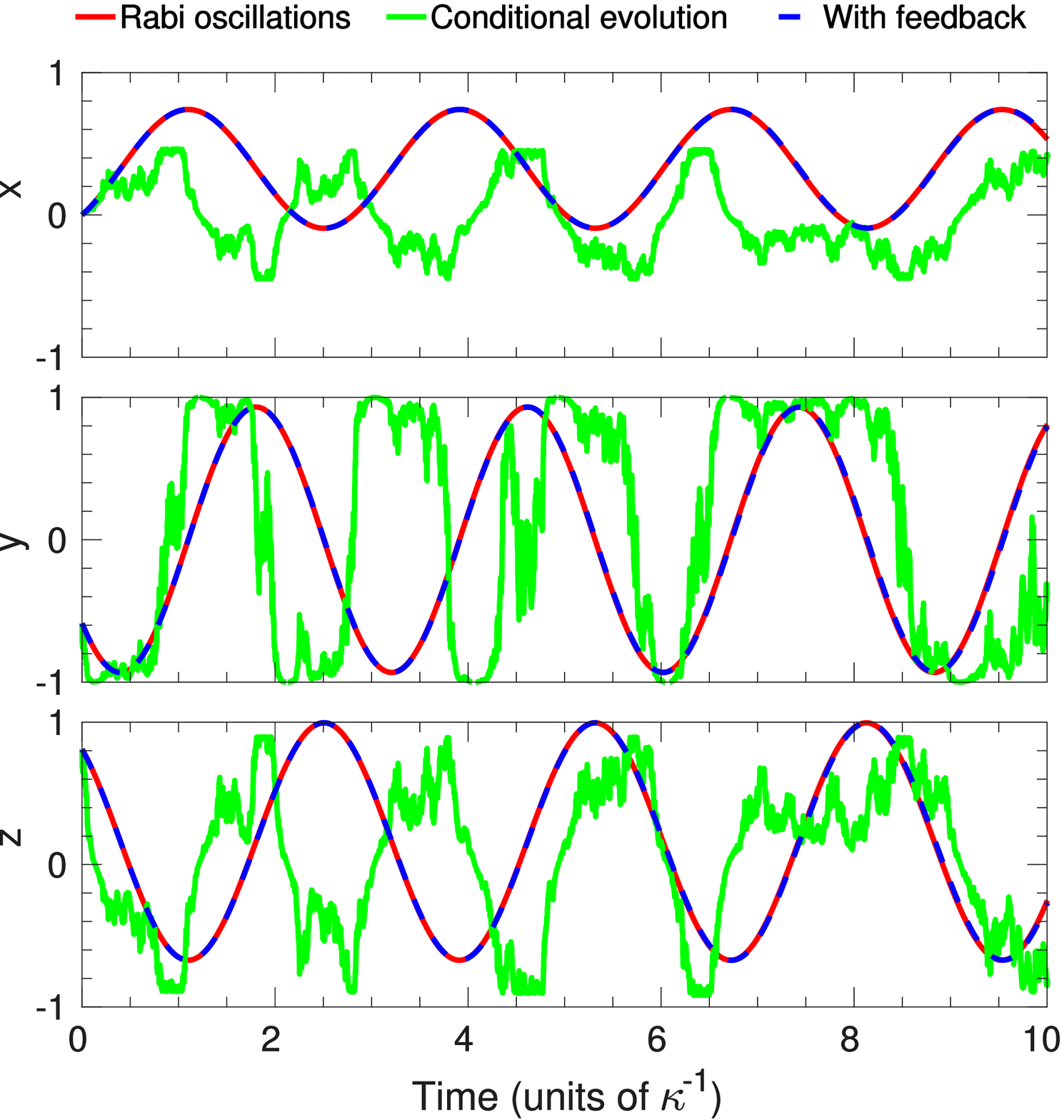}}
\\
\subfloat[][\label{subfig:3b}]{\includegraphics[width=0.45\textwidth]{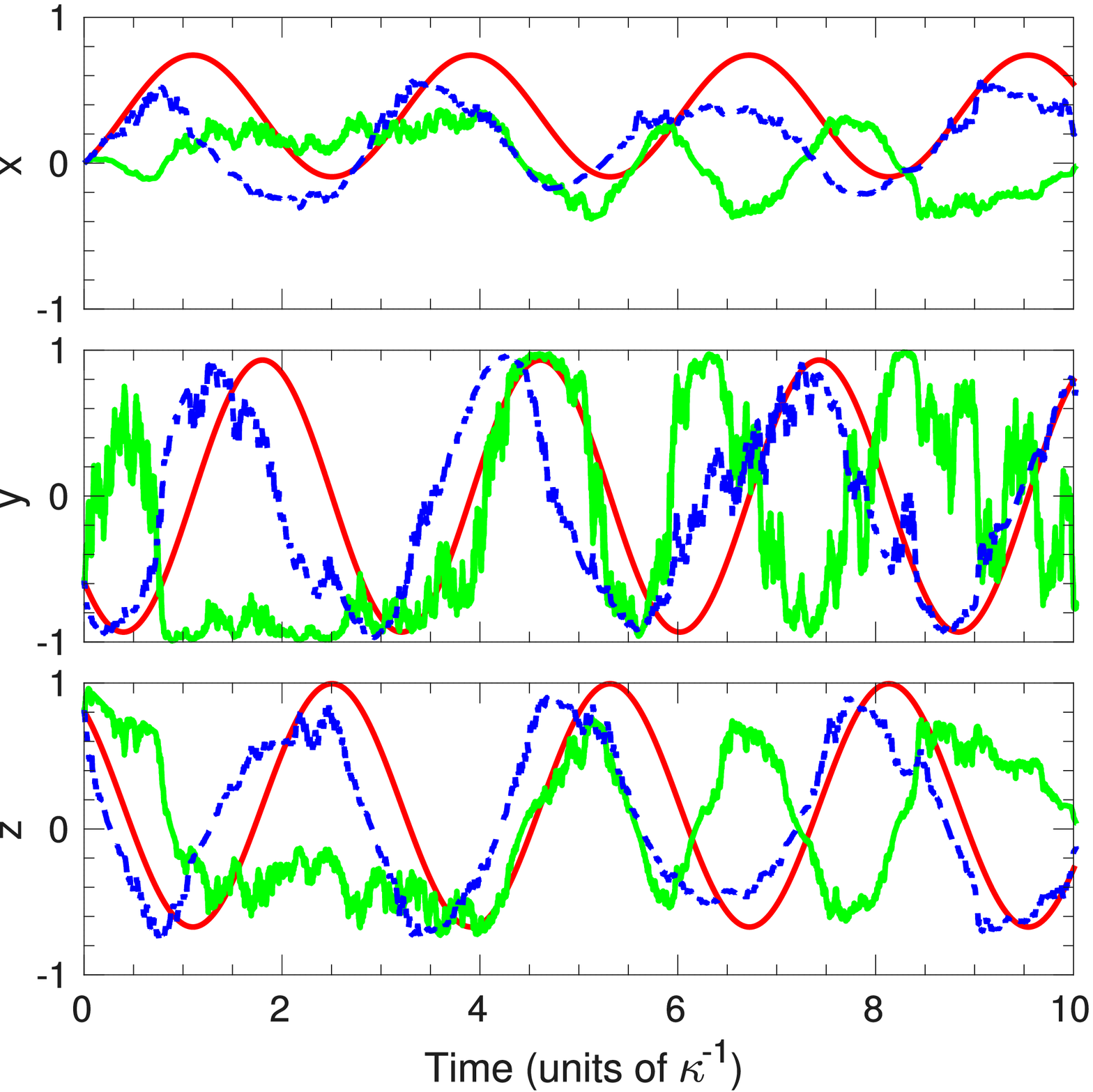}}
\caption{(Color online) Dynamical evolution of the three Bloch components of a two-level system subject to coherent Rabi oscillations (red curves).
Monitoring of the state by continuous probing of the $\sigma_y$ observable disturbs the evolution (noisy, green curve), but the system can be stabilized by feedback using the parameters \eqref{eq:feedbackParameters} calculated from the expected Bloch components of the unmonitored system (dashed, blue curve). The Rabi frequency is $\Omega = 2\kappa$ and the laser-atom detuning $\delta = 1\kappa$.
For unit detection efficiency ($\eta = 1$) in (a), the feedback completely eliminates the stochastic  back-action while with imperfect detection ($\eta=0.8$) in (b) this is not possible.}
\label{fig:example}
\end{figure}

To illustrate the accomplishments of the feedback protocol, we show in Fig.~\ref{fig:example} how the procedure can be used to stabilize monitored Rabi oscillations against measurement back-action. For unit detector efficiency in part (a), we see perfect restoration of the Rabi oscillations, while a reduced detector efficiency $\eta=0.8$ in part (b) of the figure prevents full recovery of the coherent dynamics.

\begin{figure}[h!]
\begin{center}
\includegraphics[width = 0.9\columnwidth]{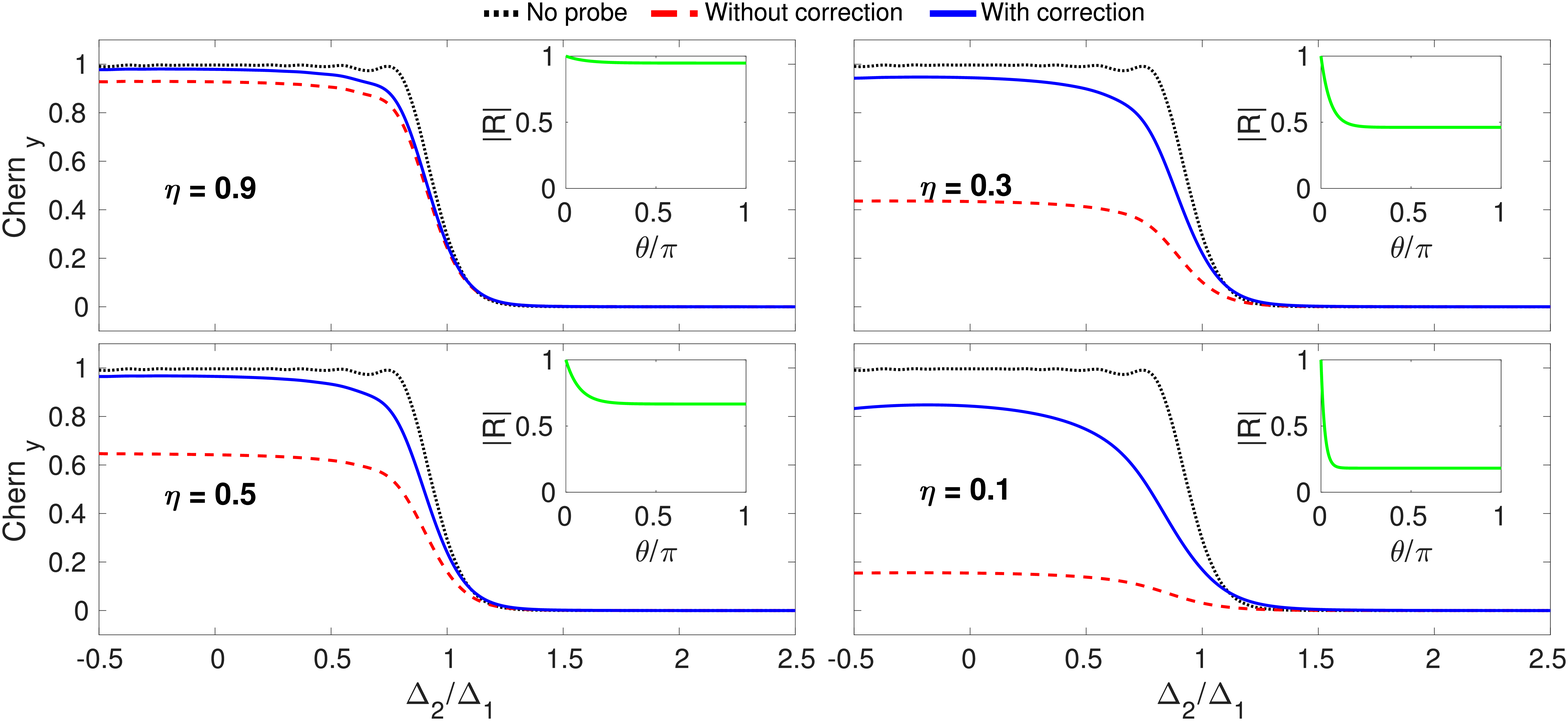}
\end{center}
\caption{(Color online) The mean value of the Chern number for different values of $\Delta_2/\Delta_1$ and for different values of the detection efficiency $\eta$.  Curves are shown for the standard expression for the Chern number, Eq.~(8) of the main text (dashed, red) and for the corrected expression, Eq.~\eqref{eq:correction} (full, blue), taking into account the reduced purity of the quantum state when subject to imperfect detection. The black, dotted curve shows, as a reference, the results in the absence of probing. The insets depict the length of the Bloch vector as a function of the sweep parameter $\theta = vt$.
Results are shown for $\Omega_1 = \frac{1}{3}\Delta_1$ and the optimized parameters (see main text); $\Delta_1/2\pi = 16.1\,M$Hz, $t_q = 0.96\,\mu$s and $\kappa/2\pi = 0.37\,M$Hz.
\label{F4abcd}}
\end{figure}

The average evolution of a system subject to continuous probing and the compensating feedback obeys an
 \emph{unconditional} master equation,
\begin{align}\label{eq:meUcFeedback}
\begin{split}
d\rho &= -i[H(t),\rho] dt-i[H_\mathrm{D}(t),\rho]dt -i[\tilde{H}_\mathrm{V}(t),\sigma_y\rho+\rho \sigma_y]dt
\\
&\quad+
\kappa\mathcal{D}[\sigma_y]\rho dt+(\kappa\eta)^{-1}\mathcal{D}[\tilde{H}_\mathrm{V}(t)]\rho dt.
\end{split}
\end{align}
With the feedback parameters given in Eqs.~\ref{eq:feedbackParameters} designed according to the Bloch components of the adiabatic eigenstate, the second term vanishes, but due to the slightly imperfect feedback, the latter three terms describe an effective attraction of the quasi-adiabatic state towards the instantaneous adiabatic eigenstate. The resulting
modifications to the Chern number estimate from the measurement current are investigated in the main text.

\section{Finite detector efficiency and a simple correction}
\label{sec:AppendixB}
As the  cancellation of the measurement back-action is crucial to achieve the detection of the Chern number by probing the qubit during a few quenches, we shall briefly explore the effect of imperfect detection for our protocol.
A finite detector efficiency $\eta$ imparts an effective non-monitored dissipation process with rate $\kappa(1-\eta)$. As shown in Ref.~\cite{2016LHenriet}, this is accompanied by a shortening of the Bloch vector over the duration of the Hamiltonian sweep and a corresponding reduction of the value of the estimated Chern number from the integral (8) of the main text.

We illustrate this effect in Fig.~\ref{F4abcd} for a system subject to our feedback by solving Eq.~\eqref{eq:meUcFeedback} for different values of the detector efficiency.
The smaller values of $\eta$ yield a larger reduction in the measurement results for the Chern number.
It is evident that the ability of the feedback to (partially) restore the undisturbed evolution of the qubit is reduced under substantial imperfections in the detection.

The state dependent control parameters of the feedback Eqs. (\ref{eq:feedbackParameters}) were derived to stabilize the instantaneous state assuming perfect detection. In the case of imperfect detection, it must be noted that the same parameters do not maximize the purity and hence the length of the Bloch vector.
An optimal scheme in the case of finite efficiency heterodyne detection is derived in the Supplemental Material of Ref. \cite{2016PCampagne-Ibarcq}, and it stands to reason that a similar approach could be used to optimize the feedback parameters in the present measurement setup.

We choose not to pursue this further at this point.
Instead we note that since the main contribution to the reduction in $C_y$ stems from the shortened $\sigma_y$ component of the Bloch vector,
we may apply a simple post-processing procedure and replace the estimator, Eq.~(8) of the main text, by
\begin{equation}\label{eq:correction}
C_y=-\int_{0}^{\pi}\frac{\Omega_{1}}{2v }\frac{\braket{\sigma_y}}{|R(t)|}\sin\theta d\theta,
\end{equation}
where $|R(t)|= \sqrt{x^2+y^2+x^2}$ is found by solving the unconditional master equation (\ref{eq:meUcFeedback}). We find that $|R(t)|$ is only very weakly dependent on the detuning parameters, and we display in insets of Fig.~\ref{F4abcd} $|R(t)|$ as a function of the sweep parameter $\theta=vt$ for $\Delta_2=2\Delta_1$. It quickly drops to a constant value which is maintained during the remaining quench sequence. The blue curves in the main plots show how Eq.~(\ref{eq:correction}) yields fairly good estimates of the Chern number as long as $\eta\gtrsim 0.3$.
Notice, however, that by Eq.~(10) of the main text, under finite detection efficiency, the noise is increased by a factor $1/\sqrt{\eta}$ which must be remedied by a correspondingly larger number of repetitions $N$.


\begin{thebibliography}{29}%
\makeatletter
\providecommand \@ifxundefined [1]{%
 \@ifx{#1\undefined}
}%
\providecommand \@ifnum [1]{%
 \ifnum #1\expandafter \@firstoftwo
 \else \expandafter \@secondoftwo
 \fi
}%
\providecommand \@ifx [1]{%
 \ifx #1\expandafter \@firstoftwo
 \else \expandafter \@secondoftwo
 \fi
}%
\providecommand \natexlab [1]{#1}%
\providecommand \enquote  [1]{``#1''}%
\providecommand \bibnamefont  [1]{#1}%
\providecommand \bibfnamefont [1]{#1}%
\providecommand \citenamefont [1]{#1}%
\providecommand \href@noop [0]{\@secondoftwo}%
\providecommand \href [0]{\begingroup \@sanitize@url \@href}%
\providecommand \@href[1]{\@@startlink{#1}\@@href}%
\providecommand \@@href[1]{\endgroup#1\@@endlink}%
\providecommand \@sanitize@url [0]{\catcode `\\12\catcode `\$12\catcode
  `\&12\catcode `\#12\catcode `\^12\catcode `\_12\catcode `\%12\relax}%
\providecommand \@@startlink[1]{}%
\providecommand \@@endlink[0]{}%
\providecommand \url  [0]{\begingroup\@sanitize@url \@url }%
\providecommand \@url [1]{\endgroup\@href {#1}{\urlprefix }}%
\providecommand \urlprefix  [0]{URL }%
\providecommand \Eprint [0]{\href }%
\providecommand \doibase [0]{http://dx.doi.org/}%
\providecommand \selectlanguage [0]{\@gobble}%
\providecommand \bibinfo  [0]{\@secondoftwo}%
\providecommand \bibfield  [0]{\@secondoftwo}%
\providecommand \translation [1]{[#1]}%
\providecommand \BibitemOpen [0]{}%
\providecommand \bibitemStop [0]{}%
\providecommand \bibitemNoStop [0]{.\EOS\space}%
\providecommand \EOS [0]{\spacefactor3000\relax}%
\providecommand \BibitemShut  [1]{\csname bibitem#1\endcsname}%
\let\auto@bib@innerbib\@empty
\bibitem [{\citenamefont {Kosterlitz}\ and\ \citenamefont
  {Thouless}(1973)}]{Kosterlitz1973}%
  \BibitemOpen
  \bibfield  {author} {\bibinfo {author} {\bibfnamefont {J.~M.}\ \bibnamefont
  {Kosterlitz}}\ and\ \bibinfo {author} {\bibfnamefont {D.~J.}\ \bibnamefont
  {Thouless}},\ }\bibfield  {title} {\enquote {\bibinfo {title} {{Ordering,
  metastability and phase transitions in two-dimensional systems}},}\ }\href
  {http://stacks.iop.org/0022-3719/6/i=7/a=010} {\bibfield  {journal} {\bibinfo
   {journal} {J. Phys. C}\ }\textbf {\bibinfo {volume} {6}},\ \bibinfo {pages}
  {1181} (\bibinfo {year} {1973})}\BibitemShut {NoStop}%
\bibitem [{\citenamefont {Thouless}\ \emph {et~al.}(1982)\citenamefont
  {Thouless}, \citenamefont {Kohmoto}, \citenamefont {Nightingale},\ and\
  \citenamefont {den Nijs}}]{Thouless1982}%
  \BibitemOpen
  \bibfield  {author} {\bibinfo {author} {\bibfnamefont {D.~J.}\ \bibnamefont
  {Thouless}}, \bibinfo {author} {\bibfnamefont {M.}~\bibnamefont {Kohmoto}},
  \bibinfo {author} {\bibfnamefont {M.~P.}\ \bibnamefont {Nightingale}}, \ and\
  \bibinfo {author} {\bibfnamefont {M.}~\bibnamefont {den Nijs}},\ }\bibfield
  {title} {\enquote {\bibinfo {title} {{Quantized Hall Conductance in a
  Two-Dimensional Periodic Potential}},}\ }\href {\doibase
  10.1103/PhysRevLett.49.405} {\bibfield  {journal} {\bibinfo  {journal} {Phys.
  Rev. Lett.}\ }\textbf {\bibinfo {volume} {49}},\ \bibinfo {pages} {405}
  (\bibinfo {year} {1982})}\BibitemShut {NoStop}%
\bibitem [{\citenamefont {Haldane}(1983)}]{Haldane1983}%
  \BibitemOpen
  \bibfield  {author} {\bibinfo {author} {\bibfnamefont {F.~D.~M.}\
  \bibnamefont {Haldane}},\ }\bibfield  {title} {\enquote {\bibinfo {title}
  {{Nonlinear Field Theory of Large-Spin Heisenberg Antiferromagnets:
  Semiclassically Quantized Solitons of the One-Dimensional Easy-Axis N\'eel
  State}},}\ }\href {\doibase 10.1103/PhysRevLett.50.1153} {\bibfield
  {journal} {\bibinfo  {journal} {Phys. Rev. Lett.}\ }\textbf {\bibinfo
  {volume} {50}},\ \bibinfo {pages} {1153} (\bibinfo {year}
  {1983})}\BibitemShut {NoStop}%
\bibitem [{\citenamefont {Niu}\ \emph {et~al.}(1985)\citenamefont {Niu},
  \citenamefont {Thouless},\ and\ \citenamefont {Wu}}]{QNiu1985}%
  \BibitemOpen
  \bibfield  {author} {\bibinfo {author} {\bibfnamefont {Q.}~\bibnamefont
  {Niu}}, \bibinfo {author} {\bibfnamefont {D.~J.}\ \bibnamefont {Thouless}}, \
  and\ \bibinfo {author} {\bibfnamefont {Y.-S.}\ \bibnamefont {Wu}},\
  }\bibfield  {title} {\enquote {\bibinfo {title} {{Quantized Hall conductance
  as a topological invariant}},}\ }\href {\doibase 10.1103/PhysRevB.31.3372}
  {\bibfield  {journal} {\bibinfo  {journal} {Phys. Rev. B}\ }\textbf {\bibinfo
  {volume} {31}},\ \bibinfo {pages} {3372} (\bibinfo {year}
  {1985})}\BibitemShut {NoStop}%
\bibitem [{\citenamefont {Fruchart}\ and\ \citenamefont
  {Carpentier}(2013)}]{Fruchart2013}%
  \BibitemOpen
  \bibfield  {author} {\bibinfo {author} {\bibfnamefont {M.}~\bibnamefont
  {Fruchart}}\ and\ \bibinfo {author} {\bibfnamefont {D.}~\bibnamefont
  {Carpentier}},\ }\bibfield  {title} {\enquote {\bibinfo {title} {{An
  introduction to topological insulators}},}\ }\href {\doibase
  http://dx.doi.org/10.1016/j.crhy.2013.09.013} {\bibfield  {journal} {\bibinfo
   {journal} {C. R. Phys}\ }\textbf {\bibinfo {volume} {14}},\ \bibinfo {pages}
  {779} (\bibinfo {year} {2013})}\BibitemShut {NoStop}%
\bibitem [{\citenamefont {Gritsev}\ and\ \citenamefont
  {Polkovnikov}(2012)}]{Gritsev2012}%
  \BibitemOpen
  \bibfield  {author} {\bibinfo {author} {\bibfnamefont {V.}~\bibnamefont
  {Gritsev}}\ and\ \bibinfo {author} {\bibfnamefont {A.}~\bibnamefont
  {Polkovnikov}},\ }\bibfield  {title} {\enquote {\bibinfo {title} {{Dynamical
  quantum Hall effect in the parameter space}},}\ }\href {\doibase
  10.1073/pnas.1116693109} {\bibfield  {journal} {\bibinfo  {journal} {Proc.
  Natl. Acad. Sci. U.S.A.}\ }\textbf {\bibinfo {volume} {109}},\ \bibinfo
  {pages} {6457} (\bibinfo {year} {2012})}\BibitemShut {NoStop}%
\bibitem [{\citenamefont {Berry}(1985)}]{MVBerry1985}%
  \BibitemOpen
  \bibfield  {author} {\bibinfo {author} {\bibfnamefont {M.~V.}\ \bibnamefont
  {Berry}},\ }\bibfield  {title} {\enquote {\bibinfo {title} {{Classical
  adiabatic angles and quantal adiabatic phase}},}\ }\href
  {http://stacks.iop.org/0305-4470/18/i=1/a=012} {\bibfield  {journal}
  {\bibinfo  {journal} {J. Phys. A}\ }\textbf {\bibinfo {volume} {18}},\
  \bibinfo {pages} {15} (\bibinfo {year} {1985})}\BibitemShut {NoStop}%
\bibitem [{\citenamefont {Chern}(1944)}]{SSChern1944}%
  \BibitemOpen
  \bibfield  {author} {\bibinfo {author} {\bibfnamefont {S.-S.}\ \bibnamefont
  {Chern}},\ }\bibfield  {title} {\enquote {\bibinfo {title} {{A simple
  intrinsic proof of the Gauss-Bonnet formula for closed Riemannian
  manifolds}},}\ }\href {http://www.jstor.org/stable/1969302} {\bibfield
  {journal} {\bibinfo  {journal} {Ann. Math.}\ }\textbf {\bibinfo {volume}
  {45}},\ \bibinfo {pages} {747} (\bibinfo {year} {1944})}\BibitemShut
  {NoStop}%
\bibitem [{\citenamefont {von Klitzing}(1986)}]{VKlitzing1986}%
  \BibitemOpen
  \bibfield  {author} {\bibinfo {author} {\bibfnamefont {K.}~\bibnamefont {von
  Klitzing}},\ }\bibfield  {title} {\enquote {\bibinfo {title} {{The quantized
  Hall effect}},}\ }\href {\doibase 10.1103/RevModPhys.58.519} {\bibfield
  {journal} {\bibinfo  {journal} {Rev. Mod. Phys.}\ }\textbf {\bibinfo {volume}
  {58}},\ \bibinfo {pages} {519} (\bibinfo {year} {1986})}\BibitemShut
  {NoStop}%
\bibitem [{\citenamefont {Haldane}(1988)}]{Haldane1988}%
  \BibitemOpen
  \bibfield  {author} {\bibinfo {author} {\bibfnamefont {F.~D.~M.}\
  \bibnamefont {Haldane}},\ }\bibfield  {title} {\enquote {\bibinfo {title}
  {{Model for a Quantum Hall Effect without Landau Levels: Condensed-Matter
  Realization of the "Parity Anomaly"}},}\ }\href {\doibase
  10.1103/PhysRevLett.61.2015} {\bibfield  {journal} {\bibinfo  {journal}
  {Phys. Rev. Lett.}\ }\textbf {\bibinfo {volume} {61}},\ \bibinfo {pages}
  {2015} (\bibinfo {year} {1988})}\BibitemShut {NoStop}%
\bibitem [{\citenamefont {Zhang}\ \emph {et~al.}(2005)\citenamefont {Zhang},
  \citenamefont {Tan}, \citenamefont {Stormer},\ and\ \citenamefont
  {Kim}}]{Zhang2005}%
  \BibitemOpen
  \bibfield  {author} {\bibinfo {author} {\bibfnamefont {Y.}~\bibnamefont
  {Zhang}}, \bibinfo {author} {\bibfnamefont {Y.-W.}\ \bibnamefont {Tan}},
  \bibinfo {author} {\bibfnamefont {H.~L.}\ \bibnamefont {Stormer}}, \ and\
  \bibinfo {author} {\bibfnamefont {P.}~\bibnamefont {Kim}},\ }\bibfield
  {title} {\enquote {\bibinfo {title} {{Experimental observation of the quantum
  Hall effect and Berry's phase in graphene}},}\ }\href@noop {} {\bibfield
  {journal} {\bibinfo  {journal} {Nature (London)}\ }\textbf {\bibinfo {volume}
  {438}},\ \bibinfo {pages} {201} (\bibinfo {year} {2005})}\BibitemShut
  {NoStop}%
\bibitem [{\citenamefont {Nayak}\ \emph {et~al.}(2008)\citenamefont {Nayak},
  \citenamefont {Simon}, \citenamefont {Stern}, \citenamefont {Freedman},\ and\
  \citenamefont {Sankar}}]{Nayak2008}%
  \BibitemOpen
  \bibfield  {author} {\bibinfo {author} {\bibfnamefont {C.}~\bibnamefont
  {Nayak}}, \bibinfo {author} {\bibfnamefont {S.~H.}\ \bibnamefont {Simon}},
  \bibinfo {author} {\bibfnamefont {A.}~\bibnamefont {Stern}}, \bibinfo
  {author} {\bibfnamefont {M.}~\bibnamefont {Freedman}}, \ and\ \bibinfo
  {author} {\bibfnamefont {D.~S.}\ \bibnamefont {Sankar}},\ }\bibfield  {title}
  {\enquote {\bibinfo {title} {{Non-Abelian anyons and topological quantum
  computation}},}\ }\href {\doibase 10.1103/RevModPhys.80.1083} {\bibfield
  {journal} {\bibinfo  {journal} {Rev. Mod. Phys.}\ }\textbf {\bibinfo {volume}
  {80}},\ \bibinfo {pages} {1083} (\bibinfo {year} {2008})}\BibitemShut
  {NoStop}%
\bibitem{foot} We study the case of a spin-1/2 qubit in an effective magnetic
 field for which the two eigenstate manifolds yield equivalent results.
\bibitem [{\citenamefont {Schroer}\ \emph {et~al.}(2014)\citenamefont
  {Schroer}, \citenamefont {Kolodrubetz}, \citenamefont {Kindel}, \citenamefont
  {Sandberg}, \citenamefont {Gao}, \citenamefont {Vissers}, \citenamefont
  {Pappas}, \citenamefont {Polkovnikov},\ and\ \citenamefont
  {Lehnert}}]{MDSchroer2014}%
  \BibitemOpen
  \bibfield  {author} {\bibinfo {author} {\bibfnamefont {M.~D.}\ \bibnamefont
  {Schroer}}, \bibinfo {author} {\bibfnamefont {M.~H.}\ \bibnamefont
  {Kolodrubetz}}, \bibinfo {author} {\bibfnamefont {W.~F.}\ \bibnamefont
  {Kindel}}, \bibinfo {author} {\bibfnamefont {M.}~\bibnamefont {Sandberg}},
  \bibinfo {author} {\bibfnamefont {J.}~\bibnamefont {Gao}}, \bibinfo {author}
  {\bibfnamefont {M.~R.}\ \bibnamefont {Vissers}}, \bibinfo {author}
  {\bibfnamefont {D.~P.}\ \bibnamefont {Pappas}}, \bibinfo {author}
  {\bibfnamefont {Anatoli}\ \bibnamefont {Polkovnikov}}, \ and\ \bibinfo
  {author} {\bibfnamefont {K.~W.}\ \bibnamefont {Lehnert}},\ }\bibfield
  {title} {\enquote {\bibinfo {title} {{Measuring a Topological Transition in
  an Artificial Spin-$1/2$ System}},}\ }\href {\doibase
  10.1103/PhysRevLett.113.050402} {\bibfield  {journal} {\bibinfo  {journal}
  {Phys. Rev. Lett.}\ }\textbf {\bibinfo {volume} {113}},\ \bibinfo {pages}
  {050402} (\bibinfo {year} {2014})}\BibitemShut {NoStop}%
\bibitem [{\citenamefont {Caves}\ \emph {et~al.}(1980)\citenamefont {Caves},
  \citenamefont {Thorne}, \citenamefont {Drever}, \citenamefont {Sandberg},\
  and\ \citenamefont {Zimmermann}}]{CMCaves1980}%
  \BibitemOpen
  \bibfield  {author} {\bibinfo {author} {\bibfnamefont {C.~M.}\ \bibnamefont
  {Caves}}, \bibinfo {author} {\bibfnamefont {K.~S.}\ \bibnamefont {Thorne}},
  \bibinfo {author} {\bibfnamefont {W.~P.}\ \bibnamefont {Drever},
  \bibfnamefont {Ronald}}, \bibinfo {author} {\bibfnamefont {V.~D.}\
  \bibnamefont {Sandberg}}, \ and\ \bibinfo {author} {\bibfnamefont
  {M.}~\bibnamefont {Zimmermann}},\ }\bibfield  {title} {\enquote {\bibinfo
  {title} {{On the measurement of a weak classical force coupled to a
  quantum-mechanical oscillator. I. Issues of principle}},}\ }\href {\doibase
  10.1103/RevModPhys.52.341} {\bibfield  {journal} {\bibinfo  {journal} {Rev.
  Mod. Phys.}\ }\textbf {\bibinfo {volume} {52}},\ \bibinfo {pages} {341}
  (\bibinfo {year} {1980})}\BibitemShut {NoStop}%
\bibitem [{\citenamefont {Clerk}\ \emph {et~al.}(2010)\citenamefont {Clerk},
  \citenamefont {Devoret}, \citenamefont {Girvin}, \citenamefont {Marquardt},\
  and\ \citenamefont {Schoelkopf}}]{AAClerk2010}%
  \BibitemOpen
  \bibfield  {author} {\bibinfo {author} {\bibfnamefont {A.~A.}\ \bibnamefont
  {Clerk}}, \bibinfo {author} {\bibfnamefont {M.~H.}\ \bibnamefont {Devoret}},
  \bibinfo {author} {\bibfnamefont {S.~M.}\ \bibnamefont {Girvin}}, \bibinfo
  {author} {\bibfnamefont {Florian}\ \bibnamefont {Marquardt}}, \ and\ \bibinfo
  {author} {\bibfnamefont {R.~J.}\ \bibnamefont {Schoelkopf}},\ }\bibfield
  {title} {\enquote {\bibinfo {title} {{Introduction to quantum noise,
  measurement, and amplification}},}\ }\href {\doibase
  10.1103/RevModPhys.82.1155} {\bibfield  {journal} {\bibinfo  {journal} {Rev.
  Mod. Phys.}\ }\textbf {\bibinfo {volume} {82}},\ \bibinfo {pages} {1155}
  (\bibinfo {year} {2010})}\BibitemShut {NoStop}%
\bibitem [{\citenamefont {Murch}\ \emph {et~al.}(2013)\citenamefont {Murch},
  \citenamefont {Weber}, \citenamefont {Macklin},\ and\ \citenamefont
  {Siddiqi}}]{KWMurch2013}%
  \BibitemOpen
  \bibfield  {author} {\bibinfo {author} {\bibfnamefont {K.~W.}\ \bibnamefont
  {Murch}}, \bibinfo {author} {\bibfnamefont {S.~J.}\ \bibnamefont {Weber}},
  \bibinfo {author} {\bibfnamefont {C.}~\bibnamefont {Macklin}}, \ and\
  \bibinfo {author} {\bibfnamefont {I.}~\bibnamefont {Siddiqi}},\ }\bibfield
  {title} {\enquote {\bibinfo {title} {{Observing single quantum trajectories
  of a superconducting quantum bit}},}\ }\href@noop {} {\bibfield  {journal}
  {\bibinfo  {journal} {Nature}\ }\textbf {\bibinfo {volume} {502}},\ \bibinfo
  {pages} {211} (\bibinfo {year} {2013})}\BibitemShut {NoStop}%
\bibitem [{\citenamefont {Hatridge}\ \emph {et~al.}(2013)\citenamefont
  {Hatridge}, \citenamefont {Shankar}, \citenamefont {Mirrahimi}, \citenamefont
  {Schackert}, \citenamefont {Geerlings}, \citenamefont {Brecht}, \citenamefont
  {Sliwa}, \citenamefont {Abdo}, \citenamefont {Frunzio}, \citenamefont
  {Girvin}, \citenamefont {Schoelkopf},\ and\ \citenamefont
  {Devoret}}]{MHatridge2013}%
  \BibitemOpen
  \bibfield  {author} {\bibinfo {author} {\bibfnamefont {M.}~\bibnamefont
  {Hatridge}}, \bibinfo {author} {\bibfnamefont {S.}~\bibnamefont {Shankar}},
  \bibinfo {author} {\bibfnamefont {M.}~\bibnamefont {Mirrahimi}}, \bibinfo
  {author} {\bibfnamefont {F.}~\bibnamefont {Schackert}}, \bibinfo {author}
  {\bibfnamefont {K.}~\bibnamefont {Geerlings}}, \bibinfo {author}
  {\bibfnamefont {T.}~\bibnamefont {Brecht}}, \bibinfo {author} {\bibfnamefont
  {K.~M.}\ \bibnamefont {Sliwa}}, \bibinfo {author} {\bibfnamefont
  {B.}~\bibnamefont {Abdo}}, \bibinfo {author} {\bibfnamefont {L.}~\bibnamefont
  {Frunzio}}, \bibinfo {author} {\bibfnamefont {S.~M.}\ \bibnamefont {Girvin}},
  \bibinfo {author} {\bibfnamefont {R.~J.}\ \bibnamefont {Schoelkopf}}, \ and\
  \bibinfo {author} {\bibfnamefont {M.~H.}\ \bibnamefont {Devoret}},\
  }\bibfield  {title} {\enquote {\bibinfo {title} {{Quantum back-action of an
  individual variable-strength measurement}},}\ }\href@noop {} {\bibfield
  {journal} {\bibinfo  {journal} {Science}\ }\textbf {\bibinfo {volume}
  {339}},\ \bibinfo {pages} {178} (\bibinfo {year} {2013})}\BibitemShut
  {NoStop}%
\bibitem [{\citenamefont {Weber}\ \emph {et~al.}(2014)\citenamefont {Weber},
  \citenamefont {Chantasri}, \citenamefont {Dressel}, \citenamefont {Jordan},
  \citenamefont {Murch},\ and\ \citenamefont {Siddiqi}}]{SJWeber2014}%
  \BibitemOpen
  \bibfield  {author} {\bibinfo {author} {\bibfnamefont {S.~J.}\ \bibnamefont
  {Weber}}, \bibinfo {author} {\bibfnamefont {A.}~\bibnamefont {Chantasri}},
  \bibinfo {author} {\bibfnamefont {J.}~\bibnamefont {Dressel}}, \bibinfo
  {author} {\bibfnamefont {A.~N.}\ \bibnamefont {Jordan}}, \bibinfo {author}
  {\bibfnamefont {K.~W.}\ \bibnamefont {Murch}}, \ and\ \bibinfo {author}
  {\bibfnamefont {I.}~\bibnamefont {Siddiqi}},\ }\bibfield  {title} {\enquote
  {\bibinfo {title} {{Mapping the optimal route between two quantum states}},}\
  }\href@noop {} {\bibfield  {journal} {\bibinfo  {journal} {Nature}\ }\textbf
  {\bibinfo {volume} {511}},\ \bibinfo {pages} {570} (\bibinfo {year}
  {2014})}\BibitemShut {NoStop}%
\bibitem [{\citenamefont {Jacobs}(2010)}]{KJacobsStochastic2010}%
  \BibitemOpen
  \bibfield  {author} {\bibinfo {author} {\bibfnamefont {K.}~\bibnamefont
  {Jacobs}},\ }\href@noop {} {\emph {\bibinfo {title} {{Stochastic Processes
  for Physicists: Understanding Noisy Systems}}}}\ (\bibinfo  {publisher}
  {Cambridge University Press, Cambridge, UK},\ \bibinfo {year} {2010})\BibitemShut
  {NoStop}%
\bibitem [{\citenamefont {Jacobs}\ and\ \citenamefont
  {Steck}(2006)}]{KJacobs2006}%
  \BibitemOpen
  \bibfield  {author} {\bibinfo {author} {\bibfnamefont {K.}~\bibnamefont
  {Jacobs}}\ and\ \bibinfo {author} {\bibfnamefont {D.~A.}\ \bibnamefont
  {Steck}},\ }\bibfield  {title} {\enquote {\bibinfo {title} {{A
  straightforward introduction to continuous quantum measurement}},}\
  }\href@noop {} {\bibfield  {journal} {\bibinfo  {journal} {Contemp. Phys.}\
  }\textbf {\bibinfo {volume} {47}},\ \bibinfo {pages} {279} (\bibinfo {year}
  {2006})}\BibitemShut {NoStop}%
\bibitem [{\citenamefont {Wiseman}\ and\ \citenamefont
  {Milburn}(2010)}]{HMWiseman2009}%
  \BibitemOpen
  \bibfield  {author} {\bibinfo {author} {\bibfnamefont {H.~M.}\ \bibnamefont
  {Wiseman}}\ and\ \bibinfo {author} {\bibfnamefont {G.~J.}\ \bibnamefont
  {Milburn}},\ }\href@noop {} {\emph {\bibinfo {title} {{Quantum Measurement
  and Control}}}}\ (\bibinfo  {publisher} {Cambridge University Press,
  Cambridge, UK},\ \bibinfo {year} {2010})\BibitemShut {NoStop}%
\bibitem [{\citenamefont {Henriet}\ \emph {et~al.}(2017)\citenamefont
  {Henriet}, \citenamefont {Sclocchi}, \citenamefont {Orth},\ and\
  \citenamefont {LeHur}}]{LHenriet2016}%
  \BibitemOpen
  \bibfield  {author} {\bibinfo {author} {\bibfnamefont {L.}~\bibnamefont
  {Henriet}}, \bibinfo {author} {\bibfnamefont {A.}~\bibnamefont {Sclocchi}},
  \bibinfo {author} {\bibfnamefont {P.~P.}\ \bibnamefont {Orth}}, \ and\
  \bibinfo {author} {\bibfnamefont {K.}~\bibnamefont {LeHur}},\ }\bibfield
  {title} {\enquote {\bibinfo {title} {{Topology of a dissipative spin:
  Dynamical Chern number, bath-induced nonadiabaticity, and a quantum dynamo
  effect}},}\ }\href {\doibase 10.1103/PhysRevB.95.054307} {\bibfield
  {journal} {\bibinfo  {journal} {Phys. Rev. B}\ }\textbf {\bibinfo {volume}
  {95}},\ \bibinfo {pages} {054307} (\bibinfo {year} {2017})}\BibitemShut
  {NoStop}%
\bibitem [{Sup()}]{Supplemental}%
  \BibitemOpen
  \href@noop {} {\bibinfo  {journal} {See Supplemental Material at [URL will be
  inserted by publisher] for a derivation of the feedback parameters and a
  discussion of the implications of finite measurement efficiency}\
  }\BibitemShut {NoStop}%
\bibitem [{\citenamefont {Hofmann}\ \emph {et~al.}(1998)\citenamefont
  {Hofmann}, \citenamefont {Mahler},\ and\ \citenamefont
  {Hess}}]{HFHofmann1998}%
  \BibitemOpen
\bibfield  {journal} {  }\bibfield  {author} {\bibinfo {author} {\bibfnamefont
  {H.~F.}\ \bibnamefont {Hofmann}}, \bibinfo {author} {\bibfnamefont
  {G.}~\bibnamefont {Mahler}}, \ and\ \bibinfo {author} {\bibfnamefont
  {O.}~\bibnamefont {Hess}},\ }\bibfield  {title} {\enquote {\bibinfo {title}
  {{Quantum control of atomic systems by homodyne detection and feedback}},}\
  }\href {\doibase 10.1103/PhysRevA.57.4877} {\bibfield  {journal} {\bibinfo
  {journal} {Phys. Rev. A}\ }\textbf {\bibinfo {volume} {57}},\ \bibinfo
  {pages} {4877} (\bibinfo {year} {1998})}\BibitemShut {NoStop}%
\bibitem [{\citenamefont {Wang}\ and\ \citenamefont
  {Wiseman}(2001)}]{JWang2001}%
  \BibitemOpen
  \bibfield  {author} {\bibinfo {author} {\bibfnamefont {J.}~\bibnamefont
  {Wang}}\ and\ \bibinfo {author} {\bibfnamefont {H.~M.}\ \bibnamefont
  {Wiseman}},\ }\bibfield  {title} {\enquote {\bibinfo {title}
  {{Feedback-stabilization of an arbitrary pure state of a two-level atom}},}\
  }\href {\doibase 10.1103/PhysRevA.64.063810} {\bibfield  {journal} {\bibinfo
  {journal} {Phys. Rev. A}\ }\textbf {\bibinfo {volume} {64}},\ \bibinfo
  {pages} {063810} (\bibinfo {year} {2001})}\BibitemShut {NoStop}%
\bibitem [{\citenamefont {Campagne-Ibarcq}\ \emph {et~al.}(2016)\citenamefont
  {Campagne-Ibarcq}, \citenamefont {Jezouin}, \citenamefont {Cottet},
  \citenamefont {Six}, \citenamefont {Bretheau}, \citenamefont {Mallet},
  \citenamefont {Sarlette}, \citenamefont {Rouchon},\ and\ \citenamefont
  {Huard}}]{PCampagne-Ibarcq2016}%
  \BibitemOpen
  \bibfield  {author} {\bibinfo {author} {\bibfnamefont {P.}~\bibnamefont
  {Campagne-Ibarcq}}, \bibinfo {author} {\bibfnamefont {S.}~\bibnamefont
  {Jezouin}}, \bibinfo {author} {\bibfnamefont {N.}~\bibnamefont {Cottet}},
  \bibinfo {author} {\bibfnamefont {P.}~\bibnamefont {Six}}, \bibinfo {author}
  {\bibfnamefont {L.}~\bibnamefont {Bretheau}}, \bibinfo {author}
  {\bibfnamefont {F.}~\bibnamefont {Mallet}}, \bibinfo {author} {\bibfnamefont
  {A.}~\bibnamefont {Sarlette}}, \bibinfo {author} {\bibfnamefont
  {P.}~\bibnamefont {Rouchon}}, \ and\ \bibinfo {author} {\bibfnamefont
  {B.}~\bibnamefont {Huard}},\ }\bibfield  {title} {\enquote {\bibinfo {title}
  {{Using Spontaneous Emission of a Qubit as a Resource for Feedback
  Control}},}\ }\href {\doibase 10.1103/PhysRevLett.117.060502} {\bibfield
  {journal} {\bibinfo  {journal} {Phys. Rev. Lett.}\ }\textbf {\bibinfo
  {volume} {117}},\ \bibinfo {pages} {060502} (\bibinfo {year}
  {2016})}\BibitemShut {NoStop}%
\bibitem [{\citenamefont {Vijay}\ \emph {et~al.}(2012)\citenamefont {Vijay},
  \citenamefont {Macklin}, \citenamefont {Slichter}, \citenamefont {Weber},
  \citenamefont {Murch}, \citenamefont {Naik}, \citenamefont {Korotkov},\ and\
  \citenamefont {Siddiqi}}]{RVijay2012}%
  \BibitemOpen
  \bibfield  {author} {\bibinfo {author} {\bibfnamefont {R.}~\bibnamefont
  {Vijay}}, \bibinfo {author} {\bibfnamefont {C.}~\bibnamefont {Macklin}},
  \bibinfo {author} {\bibfnamefont {D.~H.}\ \bibnamefont {Slichter}}, \bibinfo
  {author} {\bibfnamefont {S.~J.}\ \bibnamefont {Weber}}, \bibinfo {author}
  {\bibfnamefont {K.~W.}\ \bibnamefont {Murch}}, \bibinfo {author}
  {\bibfnamefont {R.}~\bibnamefont {Naik}}, \bibinfo {author} {\bibfnamefont
  {A.~N.}\ \bibnamefont {Korotkov}}, \ and\ \bibinfo {author} {\bibfnamefont
  {I.}~\bibnamefont {Siddiqi}},\ }\bibfield  {title} {\enquote {\bibinfo
  {title} {{Stabilizing Rabi oscillations in a superconducting qubit using
  quantum feedback}},}\ }\href@noop {} {\bibfield  {journal} {\bibinfo
  {journal} {Nature (London)}\ }\textbf {\bibinfo {volume} {490}},\ \bibinfo
  {pages} {77} (\bibinfo {year} {2012})}\BibitemShut {NoStop}%
\bibitem [{\citenamefont {Bitter}\ and\ \citenamefont
  {Dubbers}(1987)}]{TBitter1987}%
  \BibitemOpen
  \bibfield  {author} {\bibinfo {author} {\bibfnamefont {T.}~\bibnamefont
  {Bitter}}\ and\ \bibinfo {author} {\bibfnamefont {D.}~\bibnamefont
  {Dubbers}},\ }\bibfield  {title} {\enquote {\bibinfo {title} {{Manifestation
  of Berry's topological phase in neutron spin rotation}},}\ }\href {\doibase
  10.1103/PhysRevLett.59.251} {\bibfield  {journal} {\bibinfo  {journal} {Phys.
  Rev. Lett.}\ }\textbf {\bibinfo {volume} {59}},\ \bibinfo {pages} {251}
  (\bibinfo {year} {1987})}\BibitemShut {NoStop}%
\bibitem [{\citenamefont {Suter}\ \emph {et~al.}(1988)\citenamefont {Suter},
  \citenamefont {Mueller},\ and\ \citenamefont {Pines}}]{DSuter1988}%
  \BibitemOpen
  \bibfield  {author} {\bibinfo {author} {\bibfnamefont {D.}~\bibnamefont
  {Suter}}, \bibinfo {author} {\bibfnamefont {K.~T.}\ \bibnamefont {Mueller}},
  \ and\ \bibinfo {author} {\bibfnamefont {A.}~\bibnamefont {Pines}},\
  }\bibfield  {title} {\enquote {\bibinfo {title} {{Study of the
  Aharonov-Anandan quantum phase by NMR interferometry}},}\ }\href {\doibase
  10.1103/PhysRevLett.60.1218} {\bibfield  {journal} {\bibinfo  {journal}
  {Phys. Rev. Lett.}\ }\textbf {\bibinfo {volume} {60}},\ \bibinfo {pages}
  {1218} (\bibinfo {year} {1988})}\BibitemShut {NoStop}%
\end{thebibliography}

\begin{thebibliography}{3}%
\makeatletter
\providecommand \@ifxundefined [1]{%
 \@ifx{#1\undefined}
}%
\providecommand \@ifnum [1]{%
 \ifnum #1\expandafter \@firstoftwo
 \else \expandafter \@secondoftwo
 \fi
}%
\providecommand \@ifx [1]{%
 \ifx #1\expandafter \@firstoftwo
 \else \expandafter \@secondoftwo
 \fi
}%
\providecommand \natexlab [1]{#1}%
\providecommand \enquote  [1]{``#1''}%
\providecommand \bibnamefont  [1]{#1}%
\providecommand \bibfnamefont [1]{#1}%
\providecommand \citenamefont [1]{#1}%
\providecommand \href@noop [0]{\@secondoftwo}%
\providecommand \href [0]{\begingroup \@sanitize@url \@href}%
\providecommand \@href[1]{\@@startlink{#1}\@@href}%
\providecommand \@@href[1]{\endgroup#1\@@endlink}%
\providecommand \@sanitize@url [0]{\catcode `\\12\catcode `\$12\catcode
  `\&12\catcode `\#12\catcode `\^12\catcode `\_12\catcode `\%12\relax}%
\providecommand \@@startlink[1]{}%
\providecommand \@@endlink[0]{}%
\providecommand \url  [0]{\begingroup\@sanitize@url \@url }%
\providecommand \@url [1]{\endgroup\@href {#1}{\urlprefix }}%
\providecommand \urlprefix  [0]{URL }%
\providecommand \Eprint [0]{\href }%
\providecommand \doibase [0]{http://dx.doi.org/}%
\providecommand \selectlanguage [0]{\@gobble}%
\providecommand \bibinfo  [0]{\@secondoftwo}%
\providecommand \bibfield  [0]{\@secondoftwo}%
\providecommand \translation [1]{[#1]}%
\providecommand \BibitemOpen [0]{}%
\providecommand \bibitemStop [0]{}%
\providecommand \bibitemNoStop [0]{.\EOS\space}%
\providecommand \EOS [0]{\spacefactor3000\relax}%
\providecommand \BibitemShut  [1]{\csname bibitem#1\endcsname}%
\let\auto@bib@innerbib\@empty
\bibitem [{\citenamefont {Wiseman}\ and\ \citenamefont
  {Milburn}(2010)}]{2009HMWiseman}%
  \BibitemOpen
  \bibfield  {author} {\bibinfo {author} {\bibfnamefont {H.~M.}\ \bibnamefont
  {Wiseman}}\ and\ \bibinfo {author} {\bibfnamefont {G.~J.}\ \bibnamefont
  {Milburn}},\ }\href@noop {} {\emph {\bibinfo {title} {{Quantum Measurement
  and Control}}}}\ (\bibinfo  {publisher} {Cambridge University Press,
  Cambridge},\ \bibinfo {year} {2010})\BibitemShut {NoStop}%
\bibitem [{\citenamefont {Henriet}\ \emph {et~al.}(2017)\citenamefont
  {Henriet}, \citenamefont {Sclocchi}, \citenamefont {Orth},\ and\
  \citenamefont {L.~Hur}}]{2016LHenriet}%
  \BibitemOpen
  \bibfield  {author} {\bibinfo {author} {\bibfnamefont {L.}~\bibnamefont
  {Henriet}}, \bibinfo {author} {\bibfnamefont {A.}~\bibnamefont {Sclocchi}},
  \bibinfo {author} {\bibfnamefont {P.~P.}\ \bibnamefont {Orth}}, \ and\
  \bibinfo {author} {\bibfnamefont {K.}~\bibnamefont {L.~Hur}},\ }\bibfield
  {title} {\enquote {\bibinfo {title} {{Topology of a dissipative spin:
  Dynamical Chern number, bath-induced nonadiabaticity, and a quantum dynamo
  effect}},}\ }\href {\doibase 10.1103/PhysRevB.95.054307} {\bibfield
  {journal} {\bibinfo  {journal} {Phys. Rev. B}\ }\textbf {\bibinfo {volume}
  {95}},\ \bibinfo {pages} {054307} (\bibinfo {year} {2017})}\BibitemShut
  {NoStop}%
\bibitem [{\citenamefont {Campagne-Ibarcq}\ \emph {et~al.}(2016)\citenamefont
  {Campagne-Ibarcq}, \citenamefont {Jezouin}, \citenamefont {Cottet},
  \citenamefont {Six}, \citenamefont {Bretheau}, \citenamefont {Mallet},
  \citenamefont {Sarlette}, \citenamefont {Rouchon},\ and\ \citenamefont
  {Huard}}]{2016PCampagne-Ibarcq}%
  \BibitemOpen
  \bibfield  {author} {\bibinfo {author} {\bibfnamefont {P.}~\bibnamefont
  {Campagne-Ibarcq}}, \bibinfo {author} {\bibfnamefont {S.}~\bibnamefont
  {Jezouin}}, \bibinfo {author} {\bibfnamefont {N.}~\bibnamefont {Cottet}},
  \bibinfo {author} {\bibfnamefont {P.}~\bibnamefont {Six}}, \bibinfo {author}
  {\bibfnamefont {L.}~\bibnamefont {Bretheau}}, \bibinfo {author}
  {\bibfnamefont {F.}~\bibnamefont {Mallet}}, \bibinfo {author} {\bibfnamefont
  {A.}~\bibnamefont {Sarlette}}, \bibinfo {author} {\bibfnamefont
  {P.}~\bibnamefont {Rouchon}}, \ and\ \bibinfo {author} {\bibfnamefont
  {B.}~\bibnamefont {Huard}},\ }\bibfield  {title} {\enquote {\bibinfo {title}
  {{Using Spontaneous Emission of a Qubit as a Resource for Feedback
  Control}},}\ }\href {\doibase 10.1103/PhysRevLett.117.060502} {\bibfield
  {journal} {\bibinfo  {journal} {Phys. Rev. Lett.}\ }\textbf {\bibinfo
  {volume} {117}},\ \bibinfo {pages} {060502} (\bibinfo {year}
  {2016})}\BibitemShut {NoStop}%
\end{thebibliography}

%

\end{document}